\documentclass[nofootinbib]{revtex4}
\usepackage{graphicx}
\usepackage{dcolumn}
\usepackage{amsmath,amssymb,epsfig}
\usepackage{paralist}
\usepackage{comment}
\usepackage{graphicx}
\usepackage{multirow}
\usepackage{color,soul}
\usepackage{suffix}
\usepackage{mathtools}

\renewcommand{\vec}[1]{\boldsymbol{\mathrm{#1}}}

\DeclareMathOperator{\Ai}{Ai}
\DeclareMathOperator{\Pe}{Pe}

\begin{document}

\title{Optical properties of an extended gravitational lens}

\author{Slava G. Turyshev$^{1}$, Viktor T. Toth$^2$}

\affiliation{\vskip 3pt
$^1$Jet Propulsion Laboratory, California Institute of Technology,\\
4800 Oak Grove Drive, Pasadena, CA 91109-0899, USA}

\affiliation{\vskip 3pt
$^2$Ottawa, Ontario K1N 9H5, Canada}

\date{\today}

\begin{abstract}

We continue to study the optical properties of the solar gravitational lens (SGL). The aim is prospective applications of the SGL for imaging purposes. We investigate the solution of Maxwell's equations for the electromagnetic (EM) field, obtained on the background of a static gravitational field of the Sun. We now treat the Sun as an extended body with a gravitational field that can be described using an infinite series of gravitational multipole moments. Studying the propagation of monochromatic EM waves in this extended solar gravitational field, we develop a wave-optical treatment of the SGL that allows us to study the caustics formed in an image plane in the SGL's strong interference region. We investigate the EM field in several important regions, namely i) the area in the inner part of the caustic and close to the optical axis, ii) the region outside the caustic, and iii) the region in the immediate vicinity of the caustic, especially around its cusps and folds.  We show that in the first two regions the physical behavior of the EM field may be understood using the method of stationary phase. However, in the immediate vicinity of the caustic the method of stationary phase is inadequate and a wave-optical treatment is necessary. Relying on the angular eikonal method, we develop a new approach to describe the EM field accurately in all regions, including the immediate vicinity of the caustics and especially near the cusps and folds.  The method allows us to investigate the EM field in this important region, which is characterized by rapidly oscillating behavior. Our results are new and can be used to describe gravitational lensing by realistic astrophysical objects, such as stars, spiral and elliptical galaxies.

\end{abstract}


\maketitle

\section{Introduction}

Most of the methods used to describe gravitational lensing rely on a point mass model that only accounts for the monopole component of the gravitational field of the lens (see \cite{Turyshev-Toth:2017} and references therein). It is only for these types of lenses that we may expect the appearance of the Einstein rings or arcs for pointlike, compact sources of light \cite{Liebes:1964,Deguchi-Watson:1986,Schneider-Ehlers-Falco:1992,Refsdal-Surdej:1994,Narayan-Bartelmann:1996}. However, realistic lenses very rarely have sufficient spherical symmetry for their gravitational fields to be described effectively by the monopole model.  Instead of Einstein rings or arcs, these lenses yield Einstein crosses or other, more complicated images of compact lensed objects.

In addition to gravitational monopoles, quadrupole gravitational lenses also received some attention in the literature \cite{Kovner:1987,Schneider-Ehlers-Falco:1992,Chu:2016}.  Typically, these attempts combine a spherically symmetric main lens with other contributions, so that the combined gravitational lensing potential may be approximated as a spherically symmetric system weakly perturbed by a quadrupole. It was recognized that such a potential leads to formation of caustics \cite{Blandford-Narayan:1986,Blandford-Kovner1988,An:2005}.  Most of these attempts (e.g., \cite{Schneider-etal:2004}) relied on the guidance from method of the stationary phase, which was used to understand the lensing geometry and to estimate the resulting light amplification.

It was long known that to describe gravitational lensing by a complex distribution of matter, it is necessary to go beyond the geometric optics approximation, WKB and the stationary phase methods \cite{Keller:1995}. A wave-optical treatment is needed to treat the highly oscillatory behavior observed near optical caustics \cite{Berry-Upstill:1982}. To address these concerns, we recently developed the {\em angular eikonal method} \cite{Turyshev-Toth:2021-multipoles}, which provides a solution to the problem of diffraction of electromagnetic (EM) waves in the gravitational field of an extended body. In that development, we went beyond a point mass approximation and characterize the body's internal matter distribution using an infinite set of spherical harmonics. Such a description is especially straightforward in the case of a rotating axisymmetric body \cite{Turyshev-Toth:2021-multipoles}.  This new wave-theoretical  solution allows us to study gravitational lensing in the presence of arbitrary gravitational multipole perturbations of a monopole gravitational field. The new method can describe a large class of astrophysical lenses.

In the present paper, we continue our study, using the angular eikonal method, of rotating axisymmetric lenses, the solar gravitational lens (SGL) in particular \cite{Turyshev-Toth:2017,Turyshev-Toth:2019,Turyshev-Toth:2019-fin-difract,Turyshev-Toth:2020-image,Turyshev-Toth:2020-extend,Toth-Turyshev:2020}. We place special emphasis on the caustic boundary of the point-spread function (PSF) of the lens that characterizes its impulse response. Expressed in the form of zonal harmonics, perturbations of the SGL's PSF beyond the monopole are dominated by the lowest order quadrupole moment, which projects light from a point source in the shape of a hypocycloid known as the astroid. The boundaries of this astroid and, in particular, its cusps (vertices) cannot be readily described using the language of geometric optics as such methods are divergent in this region. Our wave-theoretical description, in contrast, can be used to characterize these regions with ease. We are also able to recover previously known  approximations for these regions of interest that appeared in the literature.

This paper is organized as follows: In Section \ref{sec:dif-form} we investigate the new diffraction integral both in the inner part of the caustic and outside of it. We show that the geometric optics approximation fails to describe the  EM field at the caustic. In Section~\ref{sec:wave-opt} we develop a method to study the EM field in the most interesting regions of the caustic, namely in the vicinity of the cusp singularities and folds midway between cusps. We present a description of light diffraction in the strong interference region of the gravitational lens of the extended Sun.
In Section~\ref{sec:end} we discuss our results and the next steps in our investigation.  In Appendix~\ref{sec:appA}, we study light amplification at the optical axis of the SGL, accounting for all multipoles of the Sun's axisymmetric gravitational field.

\section{The extended solar gravitational lens}
\label{sec:dif-form}

The presence of gravitational multipoles changes the diffraction of light by a gravitational field. In Ref.~\cite{Turyshev-Toth:2021-multipoles}, for a high-frequency EM wave (i.e., neglecting terms $\propto(kr)^{-1}$) and for $r\gg r_g$, we derive the EM field in the strong interference region of the SGL near its optical axis, which is set by the direction to a particular target.

\subsection{Point-spread function of the extended SGL}

Following \cite{Turyshev-Toth:2021-multipoles}, we use a heliocentric coordinate system with its $z$-axis aligned with the wavevector $\vec k$, so that $\vec k=(0,0,1)$. We introduce a unit vector in the direction of the impact parameter, $\vec n_\xi$. We consider an image plane located at distance $z$ from the Sun, a point $\vec x$ located in the image plane and a unit vector in the direction of the solar axis of rotation $\vec s$:
\begin{eqnarray}
{\vec n}_\xi&=&(\cos\phi_\xi,\sin \phi_\xi,0),
\label{eq:note0}\\
{\vec x}&=&\rho(\cos\phi,\sin \phi,0),\\
{\vec s}&=&(\sin\beta_s\cos\phi_s,\sin\beta_s\sin\phi_s,\cos\beta_s).
\label{eq:note}
\end{eqnarray}
In this geometry, up to terms of ${\cal O}(\rho^2/z^2)$, the EM field in the image plane takes the form
{}
\begin{eqnarray}
    \left( \begin{aligned}
{E}_\rho& \\
{H}_\rho& \\
  \end{aligned} \right) =\left( \begin{aligned}
{H}_\phi& \\
-{E}_\phi& \\
  \end{aligned} \right) &=&
E_0
 \sqrt{2\pi kr_g}e^{i\sigma_0}B(\vec x)
  e^{i(kz -\omega t)}
 \left( \begin{aligned}
 \cos\phi& \\
 \sin\phi& \\
  \end{aligned} \right),
  \label{eq:DB-sol-rho}
\end{eqnarray}
with the remaining EM field components being negligibly small, $({E}_z, {H}_z)= {\cal O}({\rho}/{z})$. We used the constant $\sigma_0=-kr_g\ln kr_g/e-{\textstyle\frac{\pi}{4}}$ \cite{Turyshev-Toth:2019}. The quantity $B(\vec x)$  is the complex amplitude of the EM field given as
{}
\begin{eqnarray}
B(\vec x) &=&
\frac{1}{2\pi}\int_0^{2\pi} d\phi_\xi \exp\Big[-ik\Big(\sqrt{\frac{2r_g}{r}}\rho\cos(\phi_\xi-\phi)+
2r_g\sum_{n=2}^\infty \frac{J_n}{n} \Big(\frac{R_\odot }{\sqrt{2r_gr}}\Big)^n\sin^n\beta_s\cos[n(\phi_\xi-\phi_s)]\Big)\Big].
  \label{eq:B2}
\end{eqnarray}

The quantity $B(\vec x)$  is the complex amplitude of the EM field after it scatters on the  gravitational field of an extended lens with an axisymmetric gravitational field characterized by multipoles using zonal harmonics. If the presence of the gravitational multipoles can be neglected (i.e., by setting $J_n=0,  n\geq 2$ in (\ref{eq:B2})), the result (\ref{eq:B2}) reduces to the familiar form, $B_0(\vec x) =J_0(k\sqrt{{2r_g}/{r}}\rho)$ (see relevant discussion in Refs.~\cite{Turyshev-Toth:2017,Turyshev-Toth:2021-multipoles} and references therein), where  $J_0$ is the is the Bessel function of the first kind \cite{Abramovitz-Stegun:1965}.  Eq.~(\ref{eq:B2}) is a new diffraction integral that extends the previous wave-theoretical description of gravitational lensing phenomena to the case of an extended lens with an axisymmetric gravitational field. This result was originally obtained in  \cite{Turyshev-Toth:2021-multipoles}. It offers a powerful new tool to study gravitational lensing in the limit of weak gravitational fields, at the first post-Newtonian approximation of the general theory of relativity.

When applying these results to the SGL, we recognize the fact that the Sun is an axisymmetric rotating body with ``north-south'' symmetry. As such, its gravitational field is characterized by even zonal harmonics $J_{2n}$, with the odd zonal harmonic coefficients being zero, $J_{2n+1}=0$. The zonal harmonic coefficients for the Sun are determined using available tracking data from interplanetary spacecraft, yielding $J_2=(2.25\pm0.09)\times 10^{-7}$ \cite{Park-etal:2017},  and $J_4=-4.44\times 10^{-9}$, $J_6=-2.79\times 10^{-10}$, $J_8=1.48\times 10^{-11}$ \cite{Roxburgh:2001}. The $J_{10}$ and higher zonal harmonics will have negligible effect on the SGL's diffraction pattern, thus they can be safely ignored.

Although the integral (\ref{eq:B2}) deserves a dedicated study, our focus here is its squared norm, known as the point-spread function (PSF), which in the case of the SGL is given by
{}
\begin{eqnarray}
{\tt PSF}({\vec x})=|B({\vec x})|^2&=&
B({\vec x})B^*({\vec x}),
 \label{eq:psf}
\end{eqnarray}
with $B^*(\vec x)$ being the complex conjugate of $B({\vec x})$.

The PSF characterizes the optical properties of the SGL and its imaging capabilities. The PSF is derived from the Poynting vector that is used to characterize the momentum carried by an EM wave. To apply this approach to the SGL, we use overline and brackets to denote time averaging and ensemble averaging, and compute $S_z$ as
 {}
\begin{eqnarray}
S_z({\vec x})=\frac{c}{4\pi}\big<\overline{[{\rm Re}{\vec E}\times{\rm Re}{\vec H}]}_z\big>=\frac{c}{4\pi}E_0^2\,{2\pi kr_g}\,
\big<\overline{\big({\rm Re}\big[{ B}({\vec x})e^{i(kz-\omega t)}\big]\big)^2}\big>,
  \label{eq:S_z*6z}
\end{eqnarray}
with ${\bar S}_\rho= {\bar S}_\phi=0$ for any practical purposes \cite{Turyshev-Toth:2017,Turyshev-Toth:2021-multipoles}.

Defining light amplification as usual \cite{Turyshev-Toth:2017,Turyshev-Toth:2019,Turyshev-Toth:2020-extend}, $\mu_z({\vec x})=S_z({\vec x})/|\vec S_0({\vec x})|$, where $\vec S_0({\vec x})=({c}/{8\pi})E_0^2\, \vec k$ is the Poynting vector carried by a plane wave in a vacuum in flat spacetime, we have the light amplification of the SGL given by the following expression:
 {}
\begin{eqnarray}
\mu_z({\vec x})={2\pi kr_g}  \, {\tt PSF}({\vec x}),
  \label{eq:S_mu}
\end{eqnarray}
where {\tt PSF} is given by (\ref{eq:psf}).
Using zonal harmonic coefficients, we have extended the PSF of the SGL from that of a monopole, ${\tt PSF}_0(\vec x)=J^2_0(k\sqrt{{2r_g}/{r}}\rho)$ (as discussed in  \cite{Turyshev-Toth:2017}) to the PSF given by (\ref{eq:psf}), which now includes contributions from the axisymmetric gravitational field of the Sun.

The result (\ref{eq:psf}) determines the amplitude of the EM field on the image plane in the strong interference region of the SGL, describing light received on the image plane from a point source at infinity. The integral given by (\ref{eq:psf}) governs the diffraction and interference of light that passes by the vicinity of the Sun, and characterizes the formation of caustics that emerge in the image plane.
This integral is computable but rapidly oscillating, which makes it challenging to understand its properties and its behavior. The study of this integral and its physical implications on image formation is our main objective.

\subsection{The caustics of the solar gravitational field}
\label{sec:n-caustics}

Numerical investigations of  (\ref{eq:psf}) reveal that the PSF of the SGL produces caustics in the image plane in the strong interference region \cite{Turyshev-Toth:2021-multipoles}.  In particular, the quadrupole zonal harmonic coefficient $J_2$ produces the well-known astroid caustic\footnote{\url{https://mathworld.wolfram.com/Astroid.html}}, while other multipoles contribute in the form of hypocycloid caustics\footnote{\label{foot:hypo}\url{https://mathworld.wolfram.com/Hypocycloid.html}}.

\begin{figure}
\includegraphics{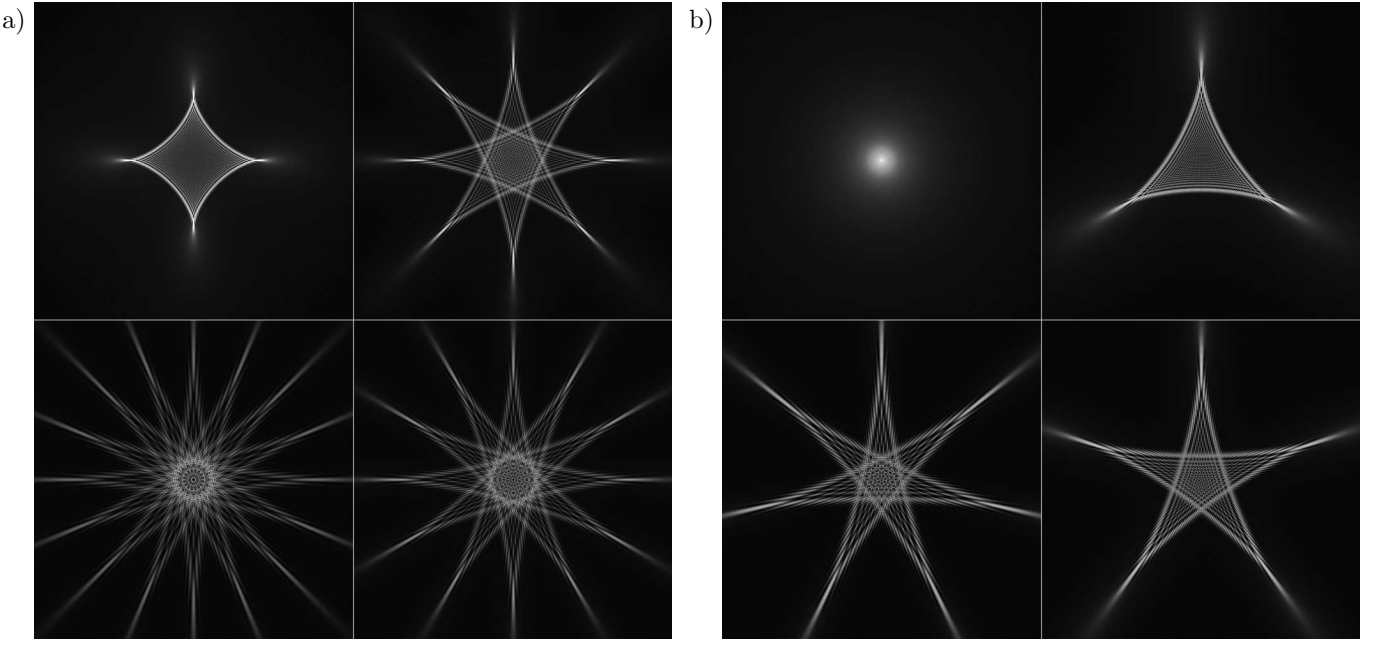}
\caption{\label{fig:caustics} Even and odd caustics representing individual contributions of the multipoles of a gravitational field to the PSF of the extended axisymmetric gravitational lens, obtained through numerical integration of ${\rm PSF}=|B({\vec x})|^2$ with $B({\vec x})$ from (\ref{eq:B2}). From top left, clockwise: a) $J_2$, $J_4$, $J_6$ and $J_8$; b) monopole, $J_3$, $J_5$ and $J_7$. The images represent a PSF calculated using $\lambda=400$~nm. For each of the $J_n$ images, the value of $J_n\sin^n\beta_s=2\times 10^{-9}$ was used with all other $J_{m\ne n}=0$, to facilitate visual comparison of their respective contributions. All other parameters are characteristic of the Sun with a $12\times 12$ meter image plane area at 650~AU. (Adapted from \cite{Turyshev-Toth:2021-multipoles}.)
}
\end{figure}

This appears to be the consequence of the complex amplitude (\ref{eq:B2}) behaving as a system of harmonic oscillators with various spatial frequencies, defined by the individual zonal harmonics. This leads to the formation of several areas of interest in the image plane. Specifically,
\begin{inparaenum}[i)]
\item In the case when $\rho$ is small, the integral is dominated by the contribution from the zonal harmonics;
\item As $\rho$ gets larger, the contribution from the monopole term and the zonal harmonics become comparable in frequency, resulting in constructive interference that manifests itself in the form of sharp contours that we recognize as the caustic boundary;
\item As $\rho$ grows further, contributions from the zonal harmonics diminish; as the monopole term reasserts its dominance, the system settles to the familiar monopole pattern \cite{Turyshev-Toth:2017} (also see discussion in Secs.~V.D.--E. of \cite{Turyshev-Toth:2021-multipoles}).
\end{inparaenum}

The curve produced by a fixed point {\it P} on the circumference of a small circle of radius $b$ rolling around the inside of a large circle of radius $a>b$ produces a hypocycloid with well-established properties \cite{Yates:1952}. Our numerical analysis shows that individual zonal harmonics in the complex amplitude of the EM field given by (\ref{eq:B2}) lead to corresponding versions of the PSF whose caustic boundaries are in the shape of appropriate hypocycloids.  To each zonal harmonic coefficient $J_n$ there corresponds a unique hypocycloid. It is natural to ask how we can recover this observed shape of the caustic boundaries directly from the integral (\ref{eq:B2}). Specifically, given a general parametric form of the equations for a caustic,
{}
\begin{eqnarray}
x&=&(a-b) \cos\phi+b\cos \Big[\Big(\frac{a-b}{b}\Big)\phi\Big],
\label{eq:caust001}\\
y&=&(a-b) \sin\phi-b\sin \Big[\Big(\frac{a-b}{b}\Big)\phi\Big],
\label{eq:caust002}
\end{eqnarray}
what are the hypocycloid radii (see footnote \ref{foot:hypo})  $a$ and $b$ corresponding to the zonal harmonic coefficient $J_n$ from (\ref{eq:B2})?

For convenience, we define
{}
\begin{eqnarray}
\alpha&=&k\sqrt\frac{2r_g}{r},~~~~\beta_n=2kr_g  \frac{J_n}{n}\Big(\frac{R_\odot }{\sqrt{2r_gr}}\Big)^n\sin^n\beta_s,
\label{eq:zerJ}
\end{eqnarray}
transforming the complex conjugate of the EM complex amplitude $B(\rho, \phi)$ from (\ref{eq:B2}) as (keeping in mind the definitions (\ref{eq:note0})--(\ref{eq:note})):
{}
\begin{eqnarray}
B^*(\rho, \phi)&=&\frac{1}{{2\pi}}\int_0^{2\pi} d\phi_\xi \exp\Big[i\Big(\alpha\rho \cos(\phi_\xi-\phi)+\sum_{n=2}^\infty \beta_n \cos[n(\phi_\xi-\phi_s)]\Big)\Big].
\label{eq:zer*1}
\end{eqnarray}

\begin{figure}
\includegraphics{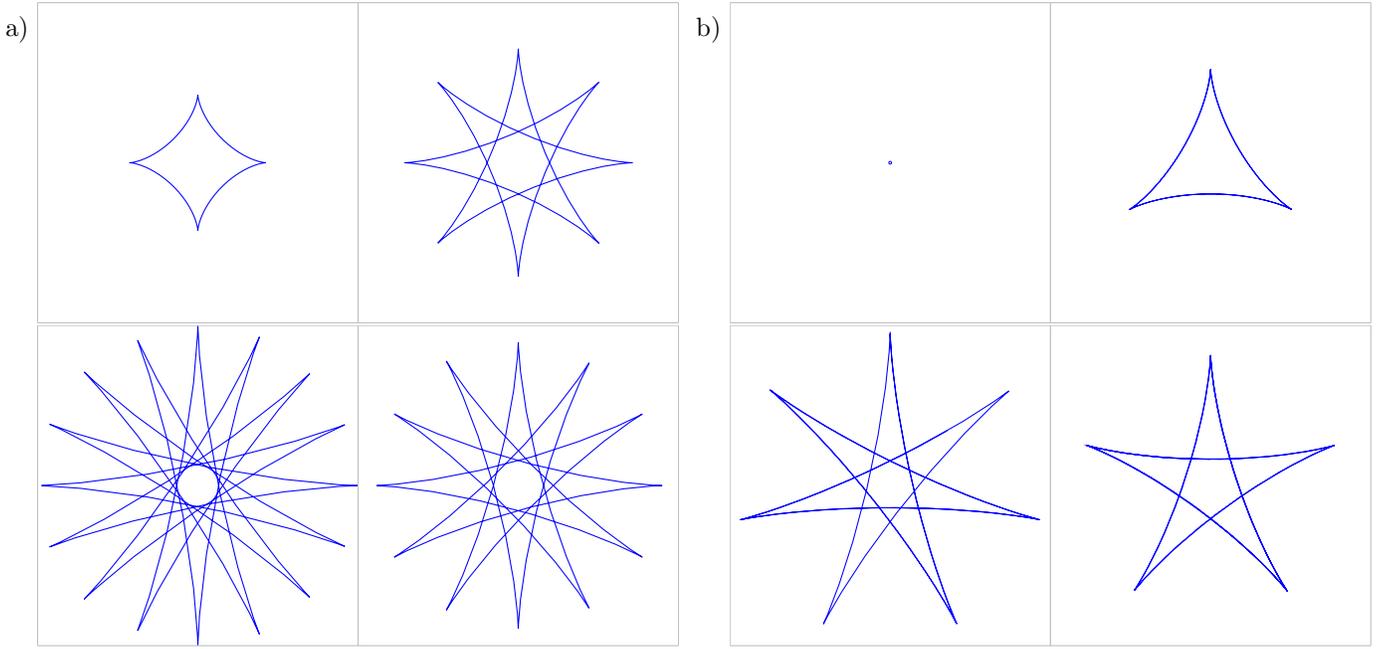}
\caption{\label{fig:caustics-model}
Caustics corresponding to Fig.~\ref{fig:caustics}, plotted using (\ref{eq:caust_h1})--(\ref{eq:caust_h2}). From top left, clockwise: a) $J_2$, $J_4$, $J_6$ and $J_8$; b) monopole (degenerate caustic), $J_3$, $J_5$ and $J_7$.
}
\end{figure}

We anticipate the caustic boundary to be characterized by a divergent expression describing light amplification. To identify such conditions, we use the method of stationary phase and consider the phase $\varphi(\rho, \phi)$ of the integral in (\ref{eq:zer*1}) that is given  as
{}
\begin{eqnarray}
\varphi(\rho, \phi)&=&\alpha\rho \cos(\phi_\xi-\phi)+\sum_{n=2}^\infty \beta_n \cos[n(\phi_\xi-\phi_s)].
\label{eq:pha0}
\end{eqnarray}
The amplification factor that is determined by the method of stationary phase is proportional $\propto \sqrt{2\pi/\varphi''}$, where $\varphi''$ is the second derivative of the phase $\varphi$ with respect to $\phi_\xi$. Therefore, the conditions that result in the vanishing of $\varphi''$  would indicate the vicinity of the caustic. Computing the needed derivative,
{}
\begin{eqnarray}
\frac{d^2\varphi(\rho, \phi)}{d\phi_\xi^2}&=&-\Big(\alpha\rho \cos(\phi_\xi-\phi)+\sum_{n=2}^\infty n^2 \beta_n \cos[n(\phi_\xi-\phi_s)]\Big),
\label{eq:pha2}
\end{eqnarray}
we see that the $n$-th caustic is formed when the amplitudes of the  terms in  (\ref{eq:pha2}) are equal. With the $\beta_n$ terms being fixed by a particular lens geometry, this condition is satisfied for specific values of $\rho$:
{}
\begin{eqnarray}
\alpha\rho = n^2\beta_n \qquad \Rightarrow \qquad
\alpha\rho_n = n^2\beta_n
 \qquad \Rightarrow \qquad
\rho_n = n^2\frac{\beta_n}{\alpha}.
\label{eq:caust}
\end{eqnarray}

We note that to form the needed caustic with the least possible number of revolutions around the angle $\phi$, the ratio between the hypocycloid radii $a$ and $b$ must be  given as
{}
\begin{eqnarray}
\frac{a}{b}=\frac{2n}{n-1}.
\label{eq:ab}
\end{eqnarray}
Each caustic is formed in $2(n-1)\pi$ revolutions around the angle $\phi$, moving counterclockwise.\footnote{Equivalently, we may form the same caustic with the ratio of $a/b=2n/(n+1)$, moving clockwise.}

The ratio (\ref{eq:ab}) and Eqs.~(\ref{eq:caust001})--(\ref{eq:caust002}), allow us to write the equations for the $n$-th caustic as
{}
\begin{eqnarray}
x_n&=&b \Big\{\Big(\frac{2n}{n-1}-1\Big) \cos\phi+\cos \Big[\Big(\frac{2n}{n-1}-1\Big)\phi\Big]\Big\},
\label{eq:caust_gg1+}\\
y_n&=&b \Big\{\Big(\frac{2n}{n-1}-1\Big) \sin\phi-\sin \Big[\Big(\frac{2n}{n-1}-1\Big)\phi\Big]\Big\}.
\label{eq:caust_gg2+}
\end{eqnarray}
Both these values have maxima at $(2n/(n-1) -1)+1=2n/(n-1)$. This prompts us to re-express (\ref{eq:caust_gg1+})--(\ref{eq:caust_gg2+}) using ${a}/b=2n/(n-1)$:
{}
\begin{eqnarray}
x_n&=&b\frac{2n}{n-1} \cdot \frac{1}{2n}  \Big\{(n+1) \cos\phi+(n-1)\cos \Big[\Big(\frac{n+1}{n-1}\Big)\phi\Big]\Big\},
\label{eq:caust_h1+}\\
y_n&=&b\frac{2n}{n-1}  \cdot\frac{1}{2n}\Big\{(n+1)\sin\phi-(n-1)\sin \Big[\Big(\frac{n+1}{n-1}\Big)\phi\Big]\Big\}.
\label{eq:caust_h2+}
\end{eqnarray}

We recognize that the maximum amplitude of the two equations above, $b[2n/(n-1)]$, is just $\rho_n$. This allows us, using (\ref{eq:caust}),  to write:
{}
\begin{eqnarray}
\rho_n=b\frac{2n}{n-1} =a=n^2\frac{\beta_n}{\alpha}
 \qquad \Rightarrow \qquad
b = n(n-1)\frac{\beta_n}{2\alpha}.
\label{eq:eqi}
\end{eqnarray}

At this point we may identify the hypocycloid radii $a$ and $b$ from (\ref{eq:caust001})--(\ref{eq:caust002}) as
{}
\begin{eqnarray}
a=\rho_n=n^2\frac{\beta_n}{\alpha}, \qquad b=n(n-1)\frac{\beta_n}{2\alpha},
\label{eq:caustd}
\end{eqnarray}
with the relationships between them at the caustic given by (\ref{eq:ab}). As a result, we may write the parametric equations that determine the shape of the $n$-th caustic as:
{}
\begin{eqnarray}
x_n&=&
n^2\frac{\beta_n}{\alpha}
\cdot
\frac{1}{2n}  \Big\{(n+1) \cos\phi+(n-1)\cos \Big[\Big(\frac{n+1}{n-1}\Big)\phi\Big]\Big\},
\label{eq:caust_h1}\\
y_n&=&
n^2\frac{\beta_n}{\alpha}
\cdot\frac{1}{2n}\Big\{(n+1)\sin\phi-(n-1)\sin \Big[\Big(\frac{n+1}{n-1}\Big)\phi\Big]\Big\},
\label{eq:caust_h2}
\end{eqnarray}
with the angle $\phi$ varying as $\phi\in [0,2(n-1)\pi]$.
Fig.~\ref{fig:caustics-model}  shows normalized caustics corresponding to the model given by (\ref{eq:caust_h1})--(\ref{eq:caust_h2}), showing precise agreement with Fig.~\ref{fig:caustics}. It is remarkable that we can now identify these caustics by reading their parameters directly off the integral (\ref{eq:zer*1}).

Therefore, using $\alpha$ and $\beta_n$ from (\ref{eq:zerJ}), the amplitude of the $n$-th caustic, $\rho_n$,  is given as
{}
\begin{eqnarray}
\rho_n = n^2\frac{\beta_n}{\alpha}= n\sqrt{2r_gr}  J_n \Big(\frac{R_\odot }{\sqrt{2r_gr}}\Big)^n\sin^n\beta_s.
\label{eq:mag}
\end{eqnarray}

With the known values of the solar multipole moments $J_2$ from \cite{Park-etal:2017},  and $J_4,J_6,J_8$ from \cite{Roxburgh:2001}, expression (\ref{eq:mag}) yields the following amplitudes of  the corresponding caustics for the largest multipole moments:
{}
\begin{eqnarray}
\rho_2 &=&2 \sqrt{2r_gr} J_2 \Big(\frac{R_\odot }{\sqrt{2r_gr}}\Big)^2\sin^2\beta_s=287.39\,{\rm m}\, \Big(\frac{650{\rm AU} }{r}\Big)^\frac{1}{2}\sin^2\beta_s,
\label{eq:mag-J2}\\
\rho_4 &=&4\sqrt{2r_gr}   J_4 \Big(\frac{R_\odot }{\sqrt{2r_gr}}\Big)^4\sin^4\beta_s=9.56\,{\rm m}\, \Big(\frac{650{\rm AU} }{r}\Big)^\frac{3}{2}\sin^4\beta_s,
\label{eq:mag-J4}\\
\rho_6 &=&6\sqrt{2r_gr}  J_6 \Big(\frac{R_\odot }{\sqrt{2r_gr}}\Big)^6\sin^6\beta_s=0.76\,{\rm m}\, \Big(\frac{650{\rm AU} }{r}\Big)^\frac{5}{2}\sin^6\beta_s,
\label{eq:mag-J6}\\
\rho_8 &=&8\sqrt{2r_gr}  J_8 \Big(\frac{R_\odot }{\sqrt{2r_gr}}\Big)^8\sin^8\beta_s=0.05\,{\rm m}\, \Big(\frac{650{\rm AU} }{r}\Big)^\frac{7}{2}\sin^8\beta_s.
\label{eq:mag-J8}
\end{eqnarray}

Even in the equatorial plane of the Sun, $\beta_s=1$, we see that only $J_2, J_4$ and $J_6$ introduce significant contributions with observable consequences if the image plane is sampled using a resolution of $\sim 1$ meter. We note that for many targets, $\beta_s<1$ , thus the magnitudes of the expressions  (\ref{eq:mag-J2})--(\ref{eq:mag-J8}) will be further suppressed by the appropriate powers of $\sin^n\beta_s$, further reducing their contributions.

\section{The EM field near the optical axis and outside the caustic}
\label{sec:stat-phase}

In Section~\ref{sec:n-caustics}, we realized that the size of the $n$-th caustic in the solar equatorial plane is directly proportional to the size of the appropriate multipole moment, $J_n$.  In the case of the Sun, the astroid  caustic set by the solar quadrupole is most prominent. Thus, it is instructive to study the PSF of the Sun by investigating the properties of the quadrupole caustic. The complex amplitude of the EM field corresponding to the quadrupole may be obtained from (\ref{eq:zer*1}) by setting $\beta_n=0, n\geq 3$, which results in the following expression:
{}
\begin{eqnarray}
B^*_2(\rho, \phi)&=&\frac{1}{{2\pi}}\int_0^{2\pi} d\phi_\xi \exp\Big[i\Big(\alpha\rho \cos(\phi_\xi-\phi)+\beta_2 \cos[2(\phi_\xi-\phi_s)]\Big)\Big].
\label{eq:B2c}
\end{eqnarray}

\begin{figure}
\includegraphics{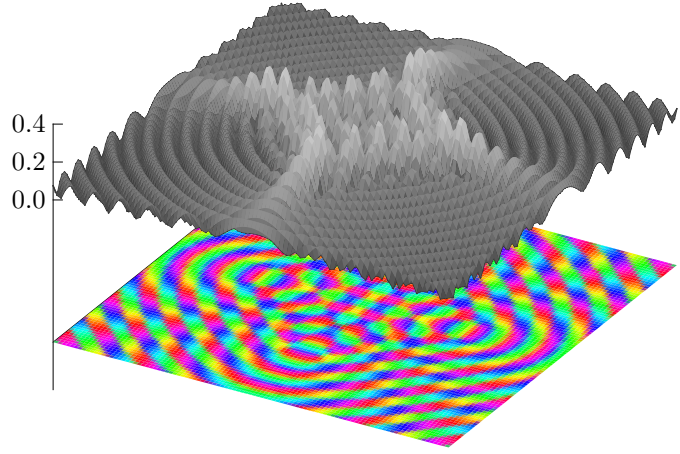}
\caption{\label{fig:caustics-B} The magnitude (top) and phase (bottom, illustrated using the red--yellow--green--cyan--blue--magenta--red cycle of colors of the rainbow corresponding to $\pi/3$ increments in phase) of the complex amplitude $B(\vec x)$ for the quadrupole (astroid) caustic. These images correspond to a $8\times 8$ meter area in the image plane of the SGL at 650~AU, with $J_2\sin^2\beta_s=2\times 10^{-9}$, $\lambda=5000$~nm.
}
\end{figure}

Fig.~\ref{fig:caustics-B}  show the magnitude and phase of the complex amplitude $B^*(\vec x)$ given by (\ref{eq:B2c}). Our objective is to investigate the behavior of this integral in various regions of the quadrupole caustic.

First, we note that $\phi_s$ simply represents a rotation of the image plane, thus we can set $\phi_s=0$ without loss of generality. We then investigate the behavior of the complex phase under the integral sign in (\ref{eq:B2c}) by writing it in the form
{}
\begin{eqnarray}
\varphi_2(\rho, \phi)=\beta_2 \Big(\cos2\overline\phi_\xi \cos2\phi-\sin2\overline\phi_\xi \sin2\phi\Big)+\alpha\rho \cos\overline\phi_\xi,
\label{eq:phB2-b}
\end{eqnarray}
where, for convenience, we introduced a new variable $\overline\phi$:
{}
\begin{eqnarray}
\overline\phi_\xi=\phi_\xi -\phi.
\label{eq:phxibar}
\end{eqnarray}

Even in the special case of $J_{n\ne 2}=0$ and $\phi_s=0$, the integral (\ref{eq:B2c}) is new, not explored in the literature. Therefore, we opt to devote our efforts to study its properties. Our goal is to investigate the behavior of this integral for specific values of $\phi$ while allowing the distance from the optical axis, $\rho$, to vary. Specifically, we will investigate two cases: $\phi=0$ and $\phi={\textstyle\frac{\pi}{4}}$.

\begin{figure}
\includegraphics{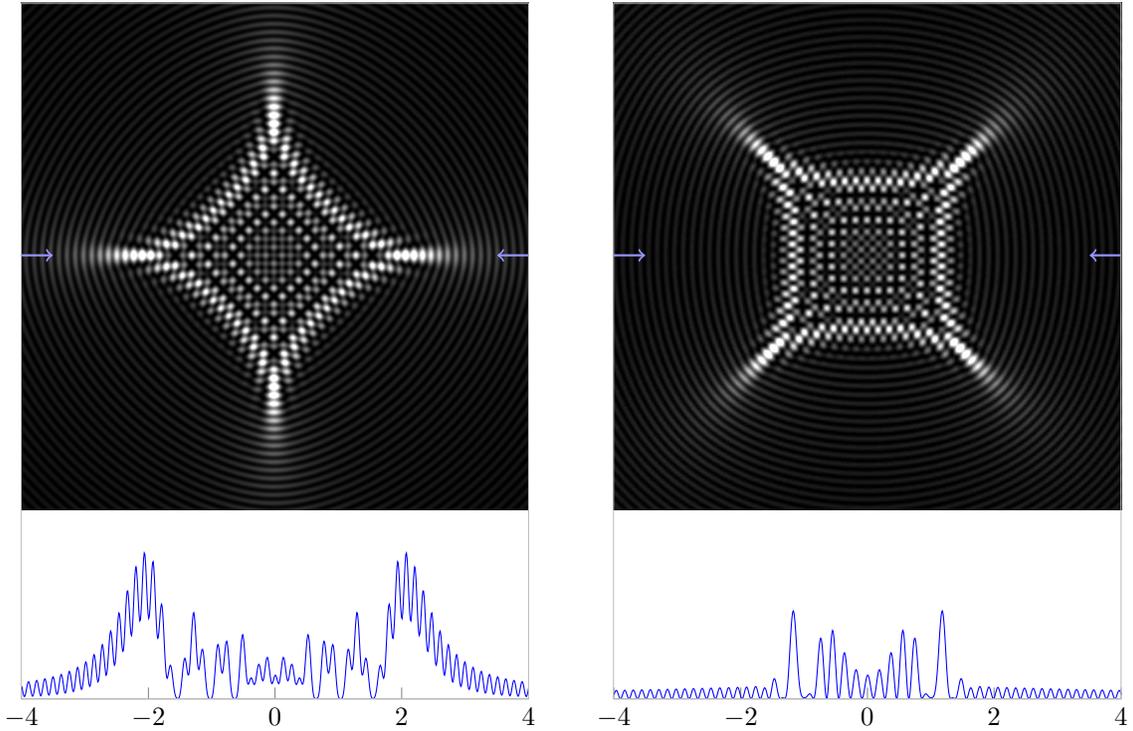}
\caption{\label{fig:caustics-cusp-all} Cross-sections of the $J_2$ (i.e., astroid) caustic. We use Eq.~(\ref{eq:B2c}) to plot the SGL PSF in a $8\times 8$ meter region in the image plane at 650 AU and for $J_2\sin^2\beta_s=2\times 10^{-9}$, $\lambda=2000$~nm.
Left: for the direction of the cusp, $\phi=0$. Right: in the  direction of the fold, $\phi={\textstyle\frac{\pi}{4}}$.}
\end{figure}

The behavior of the integral (\ref{eq:B2c}) along these directions is shown in Fig.~\ref{fig:caustics-cusp-all}. We note the high-frequency content in the inner part of the caustic, which settles down to the monopole pattern in the regions beyond the cusp and the fold for angles $\phi=0$ and $\phi={\textstyle\frac{\pi}{4}}$, correspondingly. Notice the magnitude difference between the size of the cusp and the fold.
Also, seen in Fig.~\ref{fig:caustics-cusp-all} is the highly oscillating behavior of the integral in the inner part of the casuistic. This behavior increases towards the caustic, forming sharp peaks at the cusps and folds. Moving outside the caustic, the magnitude of the oscillations sharply decreases immediately after crossing the caustic boundary. In that region, the magnitude and the frequency of the oscillations diminish, ultimately approaching the concentric pattern of the monopole PSF.

To embark on our investigation, we use (\ref{eq:caust_h1})--(\ref{eq:caust_h2}), to present parametric equations that determine the structure of the quadrupole (i.e., astroid) caustic:
 {}
\begin{eqnarray}
x_2&=&\frac{4\beta_2}{\alpha} \cdot {\textstyle\frac{1}{4} } \Big(3\cos\phi+\cos 3\phi\Big)=\frac{4\beta_2}{\alpha}\cos^3\phi,
\label{eq:q_caust_j1}\\
y_2&=&\frac{4\beta_2}{\alpha} \cdot {\textstyle\frac{1}{4} } \Big(3\sin\phi-\sin 3\phi\Big)=\frac{4\beta_2}{\alpha}\sin^3\phi.
\label{eq:q_caust_j2}
\end{eqnarray}

Therefore, by setting $\phi=0$, we can investigate the behavior of the PSF computed using (\ref{eq:B2c}) as $\rho$ increases from the optical axis (where $\rho=0$) towards the cusp region (where $\alpha\rho=4\beta_2$ or $\rho=4\beta_2/\alpha$) and beyond (for $\rho>4\beta_2/\alpha$).  Similarly, in the case of $\phi={\textstyle\frac{\pi}{4}}$, we will be able to investigate the behavior of the integral (\ref{eq:B2c})  in the valley between the cusps where the fold is formed. This, will allow us to investigate the PSF as we move from the optical axis ($\rho=0$) towards the fold region in the valley ($\rho=2\beta_2/\alpha$) and beyond (for $\rho>2\beta_2/\alpha$).

To aid with numerical evaluations of the terms involved, we estimate the magnitudes of $\alpha$ and $\beta_2$ to be
{}
\begin{eqnarray}
\alpha&=&k\sqrt\frac{2r_g}{r}=48.97\,{\rm m}^{-1}\, \Big(\frac{1\,\mu{\rm m} }{\lambda}\Big)\Big(\frac{650{\rm AU} }{r}\Big)^\frac{1}{2},
\label{eq:kbet-alp}\\
\beta_2 &=&kr_gJ_2 \Big(\frac{R_\odot }{\sqrt{2r_gr}}\Big)^2\sin^2\beta_s=3518.34\, {\rm rad}\, \Big(\frac{1\,\mu{\rm m} }{\lambda}\Big)\Big(\frac{650{\rm AU} }{r}\Big)\sin^2\beta_s,
\label{eq:kbet-J2}
\end{eqnarray}
thus we have ${4\beta_2}/{\alpha} =\rho_2=287.39\, {\rm m}\sqrt{{650\,{\rm AU} }/{r}}\sin^2\beta_s$, as given by (\ref{eq:mag-J2}).
Therefore, the size of the quadrupole caustic of the SGL is primarily determined by the angle $\beta_s$ and the heliocentric distance to the image plane.

\subsection{The PSF in the direction of the cusp}
\label{sec:GO-cusp}

When $\phi=0$, the phase (\ref{eq:phB2-b}) takes the form:
{}
\begin{eqnarray}
\varphi_2(\rho, 0)=2\beta_2 \Big(\cos\overline\phi_\xi +\frac{\alpha\rho}{4\beta_2}\Big)^2-\beta_2\Big(1+2\Big(\frac{\alpha\rho}{4\beta_2}\Big)^2\Big).
\label{eq:phB2-b0}
\end{eqnarray}
We investigate the resulting integral (\ref{eq:B2c}) using the method of stationary phase and compute
{}
\begin{eqnarray}
\frac{d\varphi_2(\rho, 0)}{d\phi_\xi}\equiv \varphi'_2(\rho, 0)&=&-4\beta_2 \sin\overline\phi_\xi\Big(\cos\overline\phi_\xi +\frac{\alpha\rho}{4\beta_2}\Big),\label{eq:phdir1}\\
\frac{d^2\varphi_2(\rho, 0)}{d\phi_\xi^2}\equiv \varphi''_2(\rho, 0)&=&-8\beta_2 \Big\{\Big(\cos\overline\phi_\xi +\frac{\alpha\rho}{16\beta_2}\Big)^2-{\frac{1}{2}}-\Big(\frac{\alpha\rho}{16\beta_2}\Big)^2\Big\}.
\label{eq:phdir2}
\end{eqnarray}

The phase is stationary when $\varphi'_2(\rho, 0)=0$, which yields four solutions
{}
\begin{eqnarray}
\overline\phi_\xi=0,\pi, \qquad {\rm and} \qquad \cos\overline\phi_\xi =-\frac{\alpha\rho}{4\beta_2},~~~
\sin\overline\phi_\xi =\pm\sqrt{1-\Big(\frac{\alpha\rho}{4\beta_2} \Big)^2}.
\label{eq:phdir1a}
\end{eqnarray}

These solutions lead to the following expressions for  $\varphi''_2(\rho, 0)$, and  $\varphi_2(\rho, 0)$:
{}
\begin{eqnarray}
\varphi''_2(\rho, 0)\big|_{\overline\phi_\xi=0}&=&-4\beta_2 \Big(1 +\frac{\alpha\rho}{4\beta_2}\Big), \qquad ~~~
\varphi_2(\rho, 0)\big|_{\overline\phi_\xi=0}=\beta_2 +\alpha\rho,
\label{eq:phdir2=0}\\
\varphi''_2(\rho, 0)\big|_{\overline\phi_\xi=\pi}&=&-4\beta_2 \Big(1 -\frac{\alpha\rho}{4\beta_2}\Big), \qquad ~~\,
\varphi_2(\rho, 0)\big|_{\overline\phi_\xi=\pi}=\beta_2 -\alpha\rho,
\label{eq:phdir2=pi}\\
\varphi''_2(\rho, 0)\big|_{\cos\overline\phi_\xi}&=&4\beta_2 \Big(1 -\Big(\frac{\alpha\rho}{4\beta_2}\Big)^2\Big), \qquad
\varphi_2(\rho, 0)\big|_{\cos\overline\phi_\xi}=-\beta_2 \Big(1 +2\Big(\frac{\alpha\rho}{4\beta_2}\Big)^2\Big).
\label{eq:phdir2=cos}
\end{eqnarray}

As we see, the second pair of solutions (\ref{eq:phdir1a}) result in the identical solutions (\ref{eq:phdir2=cos}). Thus, in the solution for $B^*$ we need to account for this solution twice.

Considering solutions (\ref{eq:phdir2=0})--(\ref{eq:phdir2=cos}), we see that they depend on the distance, $\rho$, from the optical axis. In fact, we can see that for solutions (\ref{eq:phdir2=pi}) and (\ref{eq:phdir2=cos}) the second derivatives $\varphi''_2(\rho, 0)$ change signs as $\rho$ reaches the cusp at $\rho=4\beta_2/\alpha$. The caustic boundary at the cusp marks a phase transition in the overall solution for the complex amplitude of the EM field $B^*$. This makes it necessary to consider the behavior of $B^*$ for $\rho$ separately in the following two regions:
\begin{inparaenum}[i)]
\item the inner caustic, where $0\leq\rho<4\beta_2/\alpha$ and
\item for for the outer caustic, for $\rho>4\beta_2/\alpha$.
\end{inparaenum}

There is another important observation that one can make by studying solutions (\ref{eq:phdir2=0})--(\ref{eq:phdir2=cos}), namely the second derivates given by (\ref{eq:phdir2=pi})--(\ref{eq:phdir2=cos}) are divergent at the cusp or when $\rho=4\beta_2/\alpha$. This divergence indicates the limits of the typical formulation of the method of stationary phase when dealing with the integrals with coalescing saddles in the regions when their phase highly-oscillates \cite{Stamnes-Spjelkavik:1983,Keller:1995}.  This explains that no tools used for geometric optics may be used to describe the EM field behavior in those regions \cite{Ohanian:1983}.
To deal with these regions, one needs to use different methods that we will discuss in Section~\ref{sec:wave-opt}.

Solutions (\ref{eq:phdir2=0})--(\ref{eq:phdir2=cos}) may now be used to derive in the following results for the complex amplitude by applying the method of stationary phase in the regions with a well-constrained behavior. Thus, for the region $0\leq\rho<4\beta_2/\alpha$, the solution for the complex amplitude $B^*$ takes the form:
{}
\begin{eqnarray}
B^*_{2}(\rho, 0)&=&\frac{1}{\sqrt{8\pi\beta_2\big(1-\big(\frac{\alpha\rho}{4\beta_2}\big)^2\big)}}\Big\{\sqrt{1 -\frac{\alpha\rho}{4\beta_2}}e^{i\big(\beta_2+\alpha\rho-{\textstyle\frac{\pi}{4}}\big)}+\sqrt{1 +\frac{\alpha\rho}{4\beta_2}}e^{i\big(\beta_2-\alpha\rho-{\textstyle\frac{\pi}{4}}\big)}+
2e^{-i\big(\beta_2+2\beta_2\big(\frac{\alpha\rho}{4\beta_2}\big)^2-{\textstyle\frac{\pi}{4}}\big)}\Big\}.~~~
\label{eq:Bin}
\end{eqnarray}
This results in the following expression for the ${\tt PSF}(\rho, 0)=B^*_{2}(\rho, 0)B_{2}(\rho, 0)$:
{}
\begin{eqnarray}
{\tt PSF}(\rho, 0)&=&\frac{1}{4\pi\beta_2\big(1-\big(\frac{\alpha\rho}{4\beta_2}\big)^2\big)}\Big\{3+\sqrt{1 -\big(\frac{\alpha\rho}{4\beta_2}\big)^2}\,\cos2\alpha\rho +\nonumber\\
&&+\, 2\sqrt{1 -\frac{\alpha\rho}{4\beta_2}}\,\cos\Big[2\beta_2\big(1+\frac{\alpha\rho}{4\beta_2}\big)^2-{\textstyle\frac{\pi}{2}}\Big]
+
2\sqrt{1 +\frac{\alpha\rho}{4\beta_2}}\,\cos\Big[2\beta_2\big(1-\frac{\alpha\rho}{4\beta_2}\big)^2-{\textstyle\frac{\pi}{2}}\Big]\Big\}.~~~
\label{eq:BinPSF}
\end{eqnarray}

Expression (\ref{eq:BinPSF}) allows us to evaluate the magnitude of the {\tt PSF} on the optical axis.  This can be done by setting $\rho=0$, which yields
 {}
\begin{eqnarray}
{\tt PSF}(0, 0)&=&\frac{2}{\pi\beta_2}\cos^2\big(\beta_2-{\textstyle\frac{\pi}{4}}\big).~~~
\label{eq:BinPSF0}
\end{eqnarray}
Thus, compared to the monopole PSF (which on the optical axis results in the value of ${\tt PSF}_0=J_0^2(0)=1$), the presence of the quadrupole reduces the magnitude of the PSF on the optical axis by the value given by (\ref{eq:BinPSF0}).

\begin{figure}
\includegraphics{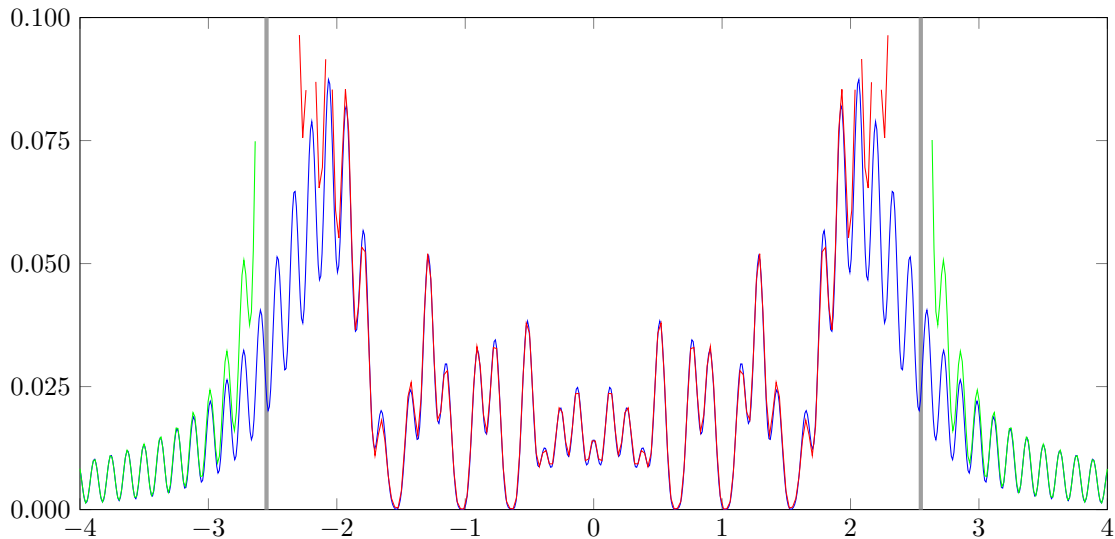}
\caption{\label{fig:caustics-cusps}
Behavior of the PSF of the SGL in the region near the optical axis in the direction towards the cusps, with parameterization corresponding to that of Fig.~\ref{fig:caustics-cusp-all} left. Blue, as given by (\ref{eq:phB2-b0}) (or, effectively, (\ref{eq:B2c}) with $\phi=0$ and $\phi_s$ absent.)  Red line shows the interior of the caustic modeled by (\ref{eq:BinPSF}). Green lines show the exterior region, as modeled by (\ref{eq:BoutPSF}). Thick gray vertical bars indicate the caustic boundary.
}
\end{figure}

Fig.~\ref{fig:caustics-cusps} presents the behavior of the PSF of the SGL in the direction towards the cusp, $\phi=0$.  The model (\ref{eq:BinPSF}) represents well both the magnitude and the frequency content of the PSF in the interior. We can see that this model also well captures the nonlinear behavior (namely, quadratic in $\alpha\rho$) of the phase. The model that was developed using the method of stationary phase works very well until the very last oscillation before reaching the peak of the cusp. Beyond the peak, the inflection point of the average amplitude marks the location where the second derivative of the two solutions (\ref{eq:phdir2=0})--(\ref{eq:phdir2=cos}) vanishes and the approximation (\ref{eq:BinPSF}) diverges. This region requires the wave-optical treatment that is developed in Sec.~\ref{sec:wave-opt}.

In Appendix~\ref{sec:appA}, we compute the value of the PSF in the presence of other multipoles.  As we recall, the spherically symmetric (monopole) PSF was given by $J^2_0(\alpha\rho)$, which in the limit $\rho\rightarrow0$ yielded 1. As seen from the result (\ref{eq:BinPSF0}) and (\ref{eq:j2-lim}), in the case of a non-negligible quadrupole  contribution, the intensity of the EM field at the optical axis is attenuated by a factor of  $({2}/{\pi\beta_2})\cos^2(\beta_2-{\textstyle\frac{\pi}{4}})$ from  (\ref{eq:BinPSF0}).

Clearly, in the case when other multipoles are also present, this value of the PSF at the optical axis is further reduced by a factor of $J^2_0(\alpha\rho)J_0^2(\beta_2)J^2_0(\beta_4)...J^2_0(\beta_{2n})$. In the case when $\beta_2, \beta_4,....\beta_{2n}$ are  large, we may use the asymptotic behavior  of the Bessel functions \cite{Abramovitz-Stegun:1965} and present the PSF at the optical axis (i.e., for $\rho=0$)
as $({2}/{\pi\beta_2})\cos^2(\beta_2-{\textstyle\frac{\pi}{4}})({2}/{\pi\beta_4})\cos^2(\beta_4-{\textstyle\frac{\pi}{4}})...({2}/{\pi\beta_{2n}})\cos^2(\beta_{2n}-{\textstyle\frac{\pi}{4}})$, where $\beta_{2n}$ are multipole terms defined similar to that of a quadrupole in (\ref{eq:zerJ=}). Depending on the value of the angle $\beta_s$, which controls the values of $\beta_{2n}$ (see (\ref{eq:zerJ})), light intensity on the optical axis may be significantly reduced by the multipoles with most of the light deflected within the caustic region, preferentially in the direction of the cusps.

For the region outside the cusp, where $\rho>4\beta_2/\alpha$, the solution for the complex amplitude $B^*$ takes the form
{}
\begin{eqnarray}
B^*_{2}(\rho, 0)&=&\frac{1}{\sqrt{8\pi\beta_2\big(\big(\frac{\alpha\rho}{4\beta_2}\big)^2-1\big)}}\Big\{\sqrt{\frac{\alpha\rho}{4\beta_2}-1}\,e^{i\big(\beta_2+\alpha\rho-{\textstyle\frac{\pi}{4}}\big)}+\sqrt{\frac{\alpha\rho}{4\beta_2}+1}\,e^{i\big(\beta_2-\alpha\rho+{\textstyle\frac{\pi}{4}}\big)}+
2e^{-i\big(\beta_2+2\beta_2\big(\frac{\alpha\rho}{4\beta_2}\big)^2+{\textstyle\frac{\pi}{4}}\big)}\Big\}.~~~
\label{eq:Bout}
\end{eqnarray}
This results in the following expression for the ${\tt PSF}(\rho, 0)=B^*_{2}(\rho, 0)B_{2}(\rho, 0)$:
{}
\begin{eqnarray}
{\tt PSF}(\rho, 0)&=&\frac{1}{4\pi\beta_2\big(\big(\frac{\alpha\rho}{4\beta_2}\big)^2-1\big)}\Big\{2\sqrt{\big(\frac{\alpha\rho}{4\beta_2}\big)^2-1}\,\cos^2\big[\alpha\rho-{\textstyle\frac{\pi}{4}}\big] +
\frac{\alpha\rho}{4\beta_2}-\sqrt{\big(\frac{\alpha\rho}{4\beta_2}\big)^2-1}+2\nonumber\\
&&+\, 2\sqrt{\frac{\alpha\rho}{4\beta_2}-1}\,\cos\Big[2\beta_2\big(\frac{\alpha\rho}{4\beta_2}+1\big)^2\Big]-
2\sqrt{\frac{\alpha\rho}{4\beta_2}+1}\,\sin\Big[2\beta_2\big(\frac{\alpha\rho}{4\beta_2}-1\big)^2\Big]\Big\}.~~~
\label{eq:BoutPSF}
\end{eqnarray}
Result (\ref{eq:BoutPSF}) allows us to compute the magnitude of the {\tt PSF} in the regions beyond the caustic, $\rho\gg 4\beta_2/\alpha$, that yields
 {}
\begin{eqnarray}
{\tt PSF}(\rho, 0)&=&\frac{2}{\pi\alpha\rho}\cos^2\big(\alpha\rho-{\textstyle\frac{\pi}{4}}\big).~~~
\label{eq:BoutPSF0}
\end{eqnarray}
This solution is also shown in Fig.~\ref{fig:caustics-cusps}. We can see that in the regions beyond the caustic boundary, at large distances from the optical axis, the PSF regains the properties of the monopole \cite{Turyshev-Toth:2017}.

\subsection{The PSF in the direction of the fold}
\label{sec:GO-fold}

In the direction of the caustic fold at $\phi={\textstyle\frac{\pi}{4}}$, the phase (\ref{eq:phB2-b}) takes the form
{}
\begin{eqnarray}
\varphi_2(\rho, {\textstyle\frac{\pi}{4}})=-2\beta_2 \cos\overline\phi_\xi \Big(\sin\overline\phi_\xi -\frac{\alpha\rho}{4\beta_2}\Big).
\label{eq:phB2-b0p4}
\end{eqnarray}
Again, we investigate the resulting integral (\ref{eq:B2c}) using the method of stationary phase. For that, we compute
{}
\begin{eqnarray}
\frac{d\varphi_2(\rho, {\textstyle\frac{\pi}{4}})}{d\phi_\xi}\equiv \varphi'_2(\rho, {\textstyle\frac{\pi}{4}})&=&4\beta_2 \Big\{\Big(\sin\overline\phi_\xi -\frac{\alpha\rho}{8\beta_2}\Big)^2-{\frac{1}{2}}-\Big(\frac{\alpha\rho}{8\beta_2}\Big)^2\Big\}
,\label{eq:phdir1p4}\\
\frac{d^2\varphi_2(\rho, {\textstyle\frac{\pi}{4}})}{d\phi_\xi^2}\equiv \varphi''_2(\rho, {\textstyle\frac{\pi}{4}})&=&
8\beta_2 \cos\overline\phi_\xi\Big(\sin\overline\phi_\xi -\frac{\alpha\rho}{8\beta_2}\Big).
\label{eq:phdir2p4}
\end{eqnarray}

The phase is stationary when $\varphi'_2(\rho, {\textstyle\frac{\pi}{4}})=0$, which yields
{}
\begin{eqnarray}
\sin\overline\phi_\xi =\frac{1}{2}\Big(\frac{\alpha\rho}{4\beta_2}\pm\sqrt{\Big(\frac{\alpha\rho}{4\beta_2}\Big)^2+2}\Big).
\label{eq:phdir1ap4}
\end{eqnarray}
As we see, at the optical axis, where $\rho=0$, this solution yeilds two acceptable solutions of $\sin\overline\phi_\xi =\pm{1}/{\sqrt{2}}$. Thus, both of these solutions will be used.  Using the result (\ref{eq:phdir1ap4}) we compute
{}
\begin{eqnarray}
\cos^2\overline\phi_\xi =\frac{1}{2}\frac{\sqrt{\Big(\frac{\alpha\rho}{4\beta_2}\Big)^2+2}\mp \frac{3\alpha\rho}{4\beta_2}}{\sqrt{\Big(\frac{\alpha\rho}{4\beta_2}\Big)^2+2}\mp \frac{\alpha\rho}{4\beta_2}}.
\label{eq:phdir1Cap4}
\end{eqnarray}

Results (\ref{eq:phdir1ap4}) and (\ref{eq:phdir1Cap4})  yield the following expressions for  $\sin\overline\phi_\xi $ and $\cos\overline\phi_\xi $:
{}
\begin{eqnarray}
\sin\overline\phi_\xi &=&\frac{1}{2}\Big(\frac{\alpha\rho}{4\beta_2}+\sqrt{\Big(\frac{\alpha\rho}{4\beta_2}\Big)^2+2}\Big), \qquad
\cos\overline\phi_\xi =\pm {\cal A}_-,
\label{eq:phdir2=01}\\
\sin\overline\phi_\xi &=&\frac{1}{2}\Big(\frac{\alpha\rho}{4\beta_2}-\sqrt{\Big(\frac{\alpha\rho}{4\beta_2}\Big)^2+2}\Big), \qquad
\cos\overline\phi_\xi =\pm  {\cal A}_+,
\label{eq:phdir2=02}
\end{eqnarray}
where the quantity $A_\pm$ is given as
{}
\begin{eqnarray}
 {\cal A}_\pm=\Bigg[\frac{1}{2}\frac{\sqrt{\Big(\frac{\alpha\rho}{4\beta_2}\Big)^2+2}\pm\frac{3\alpha\rho}{4\beta_2}}{\sqrt{\Big(\frac{\alpha\rho}{4\beta_2}\Big)^2+2}\pm\frac{\alpha\rho}{4\beta_2}}\Bigg]^\frac{1}{2}.
\label{eq:phdir2=024}
\end{eqnarray}

Substituting (\ref{eq:phdir2=01})--(\ref{eq:phdir2=024}) in (\ref{eq:phdir2p4}) and  (\ref{eq:phB2-b0p4}), we obtain the following four pairs of expressions for  $\varphi''_2(\rho, {\textstyle\frac{\pi}{4}})$ and  $\varphi_2(\rho, {\textstyle\frac{\pi}{4}})$:
{}
\begin{eqnarray}
\varphi''_{2+}(\rho, {\textstyle\frac{\pi}{4}})&=&\mp4\beta_2 \Big(\Big(\frac{\alpha\rho}{4\beta_2}\Big)^2+2\Big)^\frac{1}{2} {\cal A}_+, \qquad
\varphi_{2+}(\rho, {\textstyle\frac{\pi}{4}})=\pm\beta_2\Big(\sqrt{\Big(\frac{\alpha\rho}{4\beta_2}\Big)^2+2}+\frac{3\alpha\rho}{4\beta_2}\Big) {\cal A}_+,
\label{eq:phdir2=pip4}\\
\varphi''_{2-}(\rho, {\textstyle\frac{\pi}{4}})&=&\pm4\beta_2 \Big(\Big(\frac{\alpha\rho}{4\beta_2}\Big)^2+2\Big)^\frac{1}{2} {\cal A}_-, \qquad
\varphi_{2-}(\rho, {\textstyle\frac{\pi}{4}})=\mp\beta_2\Big(\sqrt{\Big(\frac{\alpha\rho}{4\beta_2}\Big)^2+2}-\frac{3\alpha\rho}{4\beta_2}\Big) {\cal A}_-.
\label{eq:phdir2=0p4}
\end{eqnarray}

As we see, the pair of solutions (\ref{eq:phdir2=0p4}) holds a clue on the critical behavior in the vicinity of the cusp. We observe that as $\rho$ increases, the quantity ${\cal A}_-$ vanishes at the fold and for the regions beyond the fold this pair of solutions become imaginary. This transition happens when
{}
\begin{eqnarray}
\sqrt{\Big(\frac{\alpha\rho}{4\beta_2}\Big)^2+2}-\frac{3\alpha\rho}{4\beta_2}=0\qquad \Rightarrow  \qquad \rho=\frac{2\beta_2}{\alpha},
\label{eq:phtrasit}
\end{eqnarray}
which occurs for $\rho=2\beta_2/\alpha$. According to (\ref{eq:q_caust_j1})--(\ref{eq:q_caust_j2}), this is exactly the position of the caustic boundary in the direction of the fold, $\phi={\textstyle\frac{\pi}{4}}$.

Considering solutions  (\ref{eq:phdir2=pip4})--(\ref{eq:phdir2=0p4}), we see that they depend on the distance, $\rho$, from the optical axis. In fact, we see that the solutions (\ref{eq:phdir2=0p4}) depend on ${\cal A}_-$, which becomes imaginary as $\rho$ reaches the fold at $\rho=2\beta_2/\alpha$. So, the caustic boundary represents a phase transition in the solution for the complex amplitude of the EM field $B^*$. As we did near the cusp in Sec.~\ref{sec:GO-cusp}, we must therefore separately consider the behavior of $B^*$ for $\rho$ in the following two regions:
\begin{inparaenum}[i)]
\item the inner caustic, where $0\leq\rho<2\beta_2/\alpha$ and
\item for the outer caustic, for $\rho>2\beta_2/\alpha$.
\end{inparaenum}

Similar to the discussion in Sec.~\ref{sec:GO-cusp}, we observe that the method of stationary phase is not applicable in the vicinity of the fold or when $\rho\rightarrow 2\beta_2/\alpha$. Investigating this region requires different approximation methods, discussed in Section~\ref{sec:wave-opt}.

In the meantime, we may now compute the complex amplitude of the EM field. For the region $0\leq\rho<2\beta_2/\alpha$, the solution for the complex amplitude $B^*$ takes the form
{}
\begin{eqnarray}
B^*_{2}(\rho, {\textstyle\frac{\pi}{4}})&=&\frac{1}{\sqrt{2\pi\beta_2\sqrt{\Big(\frac{\alpha\rho}{4\beta_2}\Big)^2+2}}}\Big\{\frac{1}{\sqrt{{\cal A}_+}} \cos\Big[\beta_2\Big(\sqrt{\Big(\frac{\alpha\rho}{4\beta_2}\Big)^2+2}+\frac{3\alpha\rho}{4\beta_2}\Big) {\cal A}_+- {\textstyle\frac{\pi}{4}}\Big]+\nonumber\\[-8pt]
&&\hskip 90pt +\,
\frac{1}{\sqrt{{\cal A}_-}} \cos\Big[\beta_2\Big(\sqrt{\Big(\frac{\alpha\rho}{4\beta_2}\Big)^2+2}-\frac{3\alpha\rho}{4\beta_2}\Big) {\cal A}_-- {\textstyle\frac{\pi}{4}}\Big]\Big\}.~~~
\label{eq:Binp4}
\end{eqnarray}
Note that at $\phi={\textstyle\frac{\pi}{4}}$, the complex amplitude $B^*$ (\ref{eq:Binp4}) is a real-valued function, yielding the following expression for the ${\tt PSF}(\rho, {\textstyle\frac{\pi}{4}})=B^2_{2}(\rho, {\textstyle\frac{\pi}{4}})$:
{}
\begin{eqnarray}
{\tt PSF}(\rho, {\textstyle\frac{\pi}{4}})&=&\frac{1}{2\pi\beta_2\sqrt{\Big(\frac{\alpha\rho}{4\beta_2}\Big)^2+2}}\Big\{\frac{1}{\sqrt{{\cal A}_+}} \cos\Big[\beta_2\Big(\sqrt{\Big(\frac{\alpha\rho}{4\beta_2}\Big)^2+2}+\frac{3\alpha\rho}{4\beta_2}\Big) {\cal A}_+- {\textstyle\frac{\pi}{4}}\Big]+\nonumber\\[-8pt]
&&\hskip 80pt +\,\frac{1}{\sqrt{{\cal A}_-}} \cos\Big[\beta_2\Big(\sqrt{\Big(\frac{\alpha\rho}{4\beta_2}\Big)^2+2}-\frac{3\alpha\rho}{4\beta_2}\Big) {\cal A}_-- {\textstyle\frac{\pi}{4}}\Big]\Big\}^2.~~~
\label{eq:BinPSFp4}
\end{eqnarray}

\begin{figure}
\includegraphics{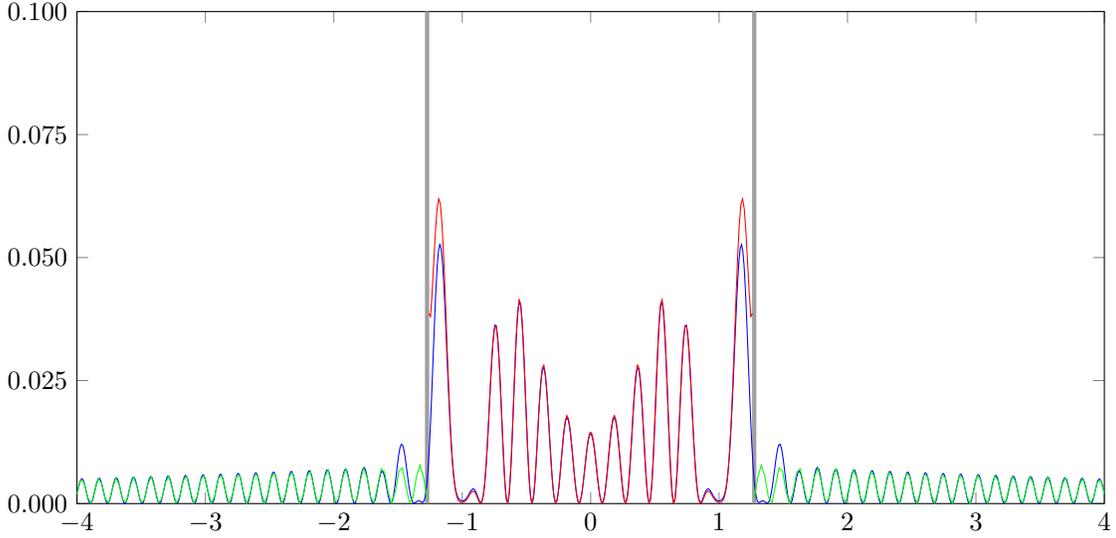}
\caption{\label{fig:caustics-fold}Behavior of the PSF of the SGL in the direction towards the fold, $\phi= {\textstyle\frac{\pi}{4}}$, as given by (\ref{eq:phB2-b0p4}) (or, effectively, (\ref{eq:B2c}) with $\phi={\textstyle\frac{\pi}{4}}$ and $\phi_s$ is absent.) The parameterization corresponds to that of Fig.~\ref{fig:caustics-cusp-all} right. The interior is well modeled by (\ref{eq:BinPSFp4}) (red), while the behavior of the exterior is well captured by (\ref{eq:BoutPSFa}) (green). Thick gray vertical bars indicate the fold regions.
}
\end{figure}

Fig.~\ref{fig:caustics-fold} presents the behavior of the PSF of the SGL
in the direction towards the fold, $\phi= {\textstyle\frac{\pi}{4}}$.
Expression (\ref{eq:BinPSFp4}) allows us to compute the magnitude of the {\tt PSF} on the optical axis.  This can be done by setting $\rho=0$, which yields
 {}
\begin{eqnarray}
{\tt PSF}(0, {\textstyle\frac{\pi}{4}})&=&\frac{2}{\pi\beta_2}\cos^2\big(\beta_2-{\textstyle\frac{\pi}{4}}\big),~~~
\label{eq:BinPSF0p4}
\end{eqnarray}
which is identical to (\ref{eq:BinPSF0}), as expected.

As we approach the caustic boundary in the direction of the fold, the the amplitude of $B^*$ determined with the method of stationary phase diverges as ${\cal A}_-$ vanishes at $\rho=2\beta_2/\alpha$. Outside the caustic boundary, $\rho>2\beta_2/\alpha$, the solution for ${\cal A}_-$ becomes imaginary. Therefore, in the region outside the caustic boundary the overall solution for the complex amplitude $B^*$ is given only by (\ref{eq:phdir2=pip4}), which results in
{}
\begin{eqnarray}
B^*_{2}(\rho, {\textstyle\frac{\pi}{4}})&=&\frac{1}{\sqrt{2\pi\beta_2\sqrt{\Big(\frac{\alpha\rho}{4\beta_2}\Big)^2+2}}}
\frac{1}{\sqrt{{\cal A}_+}} \cos\Big[\beta_2\Big(\sqrt{\Big(\frac{\alpha\rho}{4\beta_2}\Big)^2+2}+\frac{3\alpha\rho}{4\beta_2}\Big) {\cal A}_+- {\textstyle\frac{\pi}{4}}\Big].~~~
\label{eq:Boutp4}
\end{eqnarray}
This result yields in the following expression for the ${\tt PSF}(\rho, {\textstyle\frac{\pi}{4}})=B^*_{2}(\rho, {\textstyle\frac{\pi}{4}})B_{2}(\rho, {\textstyle\frac{\pi}{4}})$:
{}
\begin{eqnarray}
{\tt PSF}(\rho, {\textstyle\frac{\pi}{4}})&=&\frac{1}{2\pi\beta_2\sqrt{\Big(\frac{\alpha\rho}{4\beta_2}\Big)^2+2}}\frac{1}{{\cal A}_+} \cos^2\Big[\beta_2\Big(\sqrt{\Big(\frac{\alpha\rho}{4\beta_2}\Big)^2+2}+\frac{3\alpha\rho}{4\beta_2}\Big) {\cal A}_+- {\textstyle\frac{\pi}{4}}\Big].~~~
\label{eq:BoutPSFa}
\end{eqnarray}

Expression (\ref{eq:BoutPSFa}) allows us to compute the magnitude of the {\tt PSF} in the region beyond the caustic fold, $\rho\gg 2\beta_2/\alpha$, which yields
 {}
\begin{eqnarray}
{\tt PSF}(\rho, {\textstyle\frac{\pi}{4}})&=&\frac{2}{\pi\alpha\rho}\cos^2\big(\alpha\rho-{\textstyle\frac{\pi}{4}}\big),~~~
\label{eq:BoutPSFp4}
\end{eqnarray}
which is identical to (\ref{eq:BoutPSF0}). This is yet another confirmation that in the regions beyond the causti boundary the PSF regains the properties of that given by the monopole at large distances from the optical axis \cite{Turyshev-Toth:2017}.  Fig.~\ref{fig:caustics-fold} presents the behavior of the PSF of the SGL in the region beyond the fold in the direction $\phi= {\textstyle\frac{\pi}{4}}$, together with the approximations given by (\ref{eq:phB2-b0p4}) in the caustic interior, and by (\ref{eq:BoutPSFa}) outside. As $\rho$ gets larger, the PSF slowly decreases  ultimately matching the behavior of the monopole PSF, (\ref{eq:BoutPSFp4}).

\section{The EM field at the caustic boundary}
 \label{sec:wave-opt}

As was shown in the preceding section, the method of stationary phase fails at the caustic boundary, leading to singularities. As an outcome of this, methods relying on the geometric optics approximation are inadequate to describe the light amplification at the caustic. On the other hand, the caustic describes the region with the most light intensity. In the case of the astroid caustic formed by the quadrupole, the caustic boundary is characterized by four cusps connected by four folds, with smooth transitions between these regions.  Here we develop a wave-optical treatment of light propagation to describe these regions.

To begin, we note that there are no closed form expressions that may be used to analytically evaluate (\ref{eq:zer*1}). A way to study its behavior in the most interesting cases (i.e., the cusps, the folds between the cusps, etc.) is to develop an approximation that allows us to reduce this integral in the regions of interest to one of the canonical integrals describing cuspoid catastrophes \cite{Berry-Howles:2010,Kofler:2006}. The phases of such canonical integrals may be given in the form of $\varphi_k(t;\vec x)=t^{k+2} +\sum_{m=1}^k x_mt^m$. Special cases of such integrals involve the $k = 1$ fold catastrophe, the $k = 2$ cusp catastrophe and the $k = 3$ swallowtail catastrophe. Several such integrals may be used for our purposes.

Our first objective is to identify a coordinate transformation that may allow us to present (\ref{eq:zer*1}) in the form of the Pearcey-integral \cite{Pearcey:1946,Stamnes-Spjelkavik:1983},
{}
\begin{eqnarray}
\Pe(x,y)&=&\int_{-\infty}^{\infty}  e^{i\big(t^4+ x t^2+ y t\big)}dt,
\label{eq:Pearcey}
\end{eqnarray}
which corresponds to the case of $k=2$ and is used to describe the structure of an EM field in the neighborhood of a cusp. Similarly, we will attempt to recover a result in the form of the Airy function $\Ai(x)$,
{}
\begin{eqnarray}
\Ai(x)&=&\frac{1}{2\pi}\int_{-\infty}^{\infty}  e^{i\big(\tfrac{1}{3}t^3+ x t\big)}dt,
\label{eq:Airy}
\end{eqnarray}
which corresponds to the case of $k=1$ and is used to describe the field in the vicinity of the folds between the cusps.

To achieve these goals, we use (\ref{eq:caust_h1})--(\ref{eq:caust_h2}) for $n=2$ to define the shift of the coordinate system towards the $J_2$ caustic. We introduce a planar coordinate system at the caustic, $\widetilde{ {\vec \rho}}$, such that $\widetilde{ \rho}$ increases moving inward, towards the center of the astroid. Specifically, we introduce the coordinate transformation, $(x,y) = (x_2,y_2)+{ \widetilde{\vec \rho}}$, that is given as
  {}
\begin{eqnarray}
x&=&\frac{4\beta_2}{\alpha} \cos^3\phi -  \widetilde\rho\cos \phi,
\label{eq:caust_hx1}\\
y
&=&\frac{4\beta_2}{\alpha} \sin^3\phi - \widetilde\rho\sin \phi.
\label{eq:J2}
\end{eqnarray}

Next, as we are interested in the properties of the SGL, we consider only zonal harmonics with even $n$. Our objective is to consider the relative contributions of each of the multipoles to the intensity at the $J_2$ cusp.

Also, as we consider the PSF that is computed as ${\tt PSF}=B(x,y)B^*(x,y)$, where $B^*(x,y)$ is the complex conjugate of the amplitude $B(x,y)$, we may work with $B^*(x,y)$. With these assumptions, we transform the phase of the complex conjugate amplitude, $B^*(x,y)$, from (\ref{eq:zer*1}) as
{}
\begin{eqnarray}
\varphi(\overline \phi_\xi,\phi)&=&\beta_2 \Big(\cos2\overline\phi_\xi\cos2 \phi -\sin2\overline\phi_\xi\sin2\phi-\sin4\phi\sin\overline\phi_\xi\Big)+\Big(\beta_2\big(3+\cos4\phi\big)-\alpha\widetilde\rho\Big)\cos\overline\phi_\xi+\nonumber\\
&+&\sum_{n=2}^\infty \frac{1}{n}\beta_{2n} \Big(\cos[2n \overline \phi_\xi]\cos[2n \phi]-\sin[2n \overline \phi_\xi]\sin[2n \phi]\Big),~~~
\label{eq:ph*f1}
\end{eqnarray}
where we used  $\overline\phi_\xi=\phi_\xi-\phi$ from (\ref{eq:phxibar}).

\subsection{PSF in the vicinity of  the cusp of the astroid caustic}
\label{sec:wo-cusp}

We can now use the phase (\ref{eq:ph*f1}) to  evaluate the integral (\ref{eq:zer*1}) in the two cases of interest,  the cusp and the fold. To be specific, we will evaluate these expressions at the cusp located at $\phi=0$, and the fold defined by $\phi={\textstyle\frac{\pi}{4}}$. By setting in (\ref{eq:ph*f1})  $\phi=0$, we obtain the expression for the phase of a complex conjugate of the amplitude $B$ from (\ref{eq:zer*1}) as
{}
\begin{eqnarray}
\varphi_2(\overline \phi_\xi,0)&=&\beta_2\cos2\overline\phi_\xi +\Big(4\beta_2-\alpha\widetilde\rho\Big)\cos\overline\phi_\xi.
\label{eq:ph*0}
\end{eqnarray}

We might consider substituting this expression in (\ref{eq:zer*1})  and evaluating the resulting integral using the method of stationary phase. However, the conventional method of stationary phase fails here, as the second derivative of the phase is zero, producing divergent results.  This is the common issue with  highly oscillating integrals from the family of canonical integrals describing the cuspoid catastrophe.

Thus, we need to find other ways to approximate the integral  (\ref{eq:zer*1})  in the region of interest.  To this effect, we expand the phase (\ref{eq:ph*0}) in the vicinity of stationary points and approximate the resulting expression for the phase retaining only the leading terms with respect to powers of $\overline\phi_\xi$.  To implement the new approach, we identically present (\ref{eq:zer*1}) in a more convenient form:
{}
\begin{eqnarray}
\varphi_2(\overline \phi_\xi,0)&=&2\beta_2\Big(\cos\overline\phi_\xi +1-\frac{\alpha\widetilde\rho}{4\beta_2}\Big)^2
-\beta_2-2\beta_2\Big(1-\frac{\alpha\widetilde\rho}{4\beta_2}\Big)^2,
\label{eq:ph*0+}
\end{eqnarray}
where we separated $\overline\phi_\xi$-dependent terms and those  that are independent of this quantity.

Next, we consider the stationary points of the $J_2$ caustic only. The stationary points are given by computing the first derivative  $d\varphi_2(\overline \phi_\xi,0)/d{\phi_\xi}$. Equating it to 0, we have the equation
{}
\begin{eqnarray}
\frac{\varphi_2(\overline \phi_\xi,0)}{d\phi_\xi}&=&-4\beta_2\Big(\cos\overline\phi_\xi +1-\frac{\alpha\widetilde\rho}{4\beta_2}\Big)\sin\overline\phi_\xi=0.
\label{eq:ph-der*0+}
\end{eqnarray}

Equation (\ref{eq:ph-der*0+}) has two solutions: $\cos\overline\phi_\xi +1-{\alpha\widetilde\rho}/{4\beta_2}=0$ and $\sin\overline\phi_\xi=0$.
As we are interested in evaluating the behavior of the integral  (\ref{eq:zer*1}) as the distance $\widetilde\rho$ from the cusp changes, the second solution is of no interest. Considering the first solution, we see that for small $\widetilde\rho$, it behaves as $\cos\overline\phi_\xi =-1+{\cal O}({\alpha\widetilde\rho}/{4\beta_2})$. Thus, at the vicinity of the cusp, most of the contribution to the phase  comes from the area where $\overline\phi_\xi \approx \pi$, that is also suggested by the second solution.

We also compute the second derivative of the phase $\varphi_2(\overline \phi_\xi,0)$ as
{}
\begin{eqnarray}
\frac{d^2\varphi_2(\overline \phi_\xi,0)}{d\phi_\xi^2}&=&-8\beta_2\Big\{\Big(\cos\overline\phi_\xi +{\textstyle\frac{1}{4}}\Big(1-\frac{\alpha\widetilde\rho}{4\beta_2}\Big)\Big)^2-{\textstyle\frac{1}{2}}-\Big({\textstyle\frac{1}{4}}\Big(1-\frac{\alpha\widetilde\rho}{4\beta_2}\Big)\Big)^2\Big\}.
\label{eq:ph-der*22+}
\end{eqnarray}
Evaluating this expression for the two solutions  $\overline\phi_\xi=\{0,\pi\}$ identified above, we see that  in the case $\overline\phi_\xi=0$, the second derivative is finite everywhere in the vicinity of the cusp. However, in the case when $\overline\phi_\xi=\pi$, the second derivative (\ref{eq:ph-der*22+}) vanishes at the origin, $\widetilde\rho=0$, thus indicating the presence of the cusp.

These observations allow us to consider behavior of the phase (\ref{eq:ph*f1}) around the point $\overline\phi_\xi =\pi$, that is the direction towards the cusp. To proceed, we introduce a new variable $\widetilde\phi_\xi$:
{}
\begin{eqnarray}
\overline\phi_\xi =\pi+\widetilde\phi_\xi.
\label{eq:w-tilde}
\end{eqnarray}

At this point, we can use the complete phase at the caustic given by (\ref{eq:ph*f1}) and develop the needed approximation. For that, we substitute (\ref{eq:w-tilde}) in (\ref{eq:ph*f1}) and expand the result in a power series of the small angle  $\widetilde\phi_\xi$, retaining only leading terms:
{}
\begin{eqnarray}
\varphi(\widetilde \phi_\xi,\phi)&=&\alpha\widetilde\rho+\beta_2\Big(\sin^2 2\phi+2\sin^2\phi\cos2\phi-3\Big)-8\beta_2\sin^3\phi\cos\phi\, \widetilde\phi_\xi+\nonumber\\
&+& \Big(-{\textstyle\frac{1}{2}}\alpha\widetilde\rho +4\beta_2\sin^4\phi\Big)\widetilde\phi^2_\xi+\beta_2\sin2\phi\Big(1+{\textstyle\frac{2}{3}}\sin^2\phi\Big)\widetilde\phi^3_\xi+\nonumber\\
&+&
 \Big(-{\textstyle\frac{1}{24}}\alpha\widetilde\rho +\beta_2 \Big({\textstyle\frac{1}{2}} -\sin^2\phi-{\textstyle\frac{1}{3}}\sin^4\phi \Big)\Big)\widetilde\phi^4_\xi+\sum_{n=2}^\infty
 \beta_{2n}\cos[2n(\widetilde\phi_\xi+\pi+\phi)]+{\cal O}(\widetilde\phi^5_\xi),
\label{eq:ph*df}
\end{eqnarray}
where, for convenience, we are not yet transforming the multipolar term.

As we are interested to study the behavior of the phase in the vicinity of the cusp, we perform additional expansion of expression (\ref{eq:ph*df}) with respect to a small angle $\phi$, while treating it to be of the similar order as  $\widetilde\phi_\xi$, namely $\phi\sim\widetilde\phi_\xi$. This is the result of these manipulations:
{}
\begin{eqnarray}
\varphi(\widetilde \phi_\xi,\phi)&=& {\textstyle\frac{1}{2}} \beta_2 \cdot \widetilde\phi^4_\xi+2\beta_2 \phi\cdot \widetilde\phi^3_\xi-{\textstyle\frac{1}{2}}\alpha\widetilde\rho\cdot \widetilde\phi^2_\xi -8\beta_2\phi^3\cdot \widetilde\phi_\xi
+\sum_{n=2}^\infty
\beta_{2n}\cos[2n(\widetilde\phi_\xi+\pi+\phi)] + w_0,
\label{eq:ph*df2}
\end{eqnarray}
where the constant $w_0$ has the form $w_0=\alpha\widetilde\rho+\beta_2\big(\sin^2 2\phi+2\sin^2\phi\cos2\phi-3\big)$.

Combining the definitions used to introduce the variables  $\overline\phi_\xi$ and $\widetilde\phi_\xi$, given by (\ref{eq:phxibar}) and (\ref{eq:w-tilde}), correspondingly, we have $\widetilde\phi_\xi=\phi_\xi-\pi-\phi$. Recognizing this fact and remembering that the angle $\phi$ is small, we transform (\ref{eq:ph*df2}) as:
{}
\begin{eqnarray}
\varphi( \phi_\xi,\phi)&=&\Big({\textstyle\frac{1}{2}} \beta_2 +\sum_{n=2}^\infty {\textstyle\frac{2}{3}} n^4\beta_{2n}\Big)\big(\phi_\xi-\pi\big)^4 -\Big({\textstyle\frac{1}{2}}\alpha\widetilde\rho +3\beta_2\phi^2+\sum_{n=2}^\infty 2 n^2\beta_{2n}\Big)\big(\phi_\xi-\pi\big)^2 +\nonumber\\
&+& \Big(\alpha\widetilde\rho -4\beta_2\phi^2\Big)\phi\,\big(\phi_\xi-\pi\big)+w_0-
{\textstyle\frac{1}{2}}\alpha\widetilde\rho\phi^2+{\textstyle\frac{13}{2}}\beta_4\phi^4+\sum_{n=2}^\infty
\beta_{2n}.
\label{eq:ph*0d}
\end{eqnarray}

For the sake of simplicity, we introduce the constants $\beta^*_2$ and $\varphi_0$:
{}
\begin{eqnarray}
{\textstyle\frac{1}{2}} \beta^*_2=\Big({\textstyle\frac{1}{2}} \beta_2 +\sum_{n=2}^\infty {\textstyle\frac{2}{3}} n^4\beta_{2n}\Big),\qquad
\varphi_0=w_0-
{\textstyle\frac{1}{2}}\alpha\widetilde\rho\phi^2+{\textstyle\frac{13}{2}}\beta_4\phi^4+\sum_{n=2}^\infty
\beta_{2n}.
\end{eqnarray}
We also  introduce a rescaled distance $\overline \rho$ given by
{}
\begin{eqnarray}
\overline\rho=\widetilde\rho+ \sum_{n=2}^\infty 4 n^2\frac{\beta_{2n}}{\alpha}.
\label{eq:var_u}
\end{eqnarray}
Finally, we introduce a new variable, $u$:
{}
\begin{eqnarray}
{\textstyle\frac{1}{2}} \beta^*_2 \big(\phi_\xi-\pi\big)^4=u^4 \qquad \Rightarrow \qquad u= \big(\phi_\xi-\pi\big)\big({\textstyle\frac{1}{2}} \beta^*_2\big)^\frac{1}{4}, \qquad
\phi_\xi=\pi+u\big({\textstyle\frac{1}{2}} \beta^*_2\big)^{-{\frac{1}{4}}}.
\label{eq:var-var}
\end{eqnarray}

These notations and definitions allow us to transform the phase (\ref{eq:ph*0d}):
{}
\begin{eqnarray}
\varphi(u,\phi)&=&u^4 -\frac{(\alpha\overline\rho+6\beta_2\phi^2)}{\sqrt{2\beta^*_2}}u^2+\frac{(\alpha\overline\rho-4\beta_2\phi^2)\phi}{\big({\textstyle\frac{1}{2}}\beta^*_2\big)^\frac{1}{4}}u +\varphi_0.
\label{eq:ph*ph}
\end{eqnarray}

We may now substitute these results into the integral (\ref{eq:zer*1})  that describes the complex conjugate amplitude of the EM field. In the vicinity of the cusp  this amplitude then takes the form
{}
\begin{eqnarray}
B^*_2(\overline\rho, \phi)&=&\frac{1}{\big({\textstyle\frac{1}{2}} \beta^*_2\big)^{{\frac{1}{4}}}} \frac{e^{i\varphi_0}}{2\pi} \int_{-\infty}^{\infty} du \exp\Big[i\Big\{u^4 -\frac{(\alpha\overline\rho+6\beta_2\phi^2)}{\sqrt{2\beta^*_2}}u^2+\frac{(\alpha\overline\rho-4\beta_2\phi^2)\phi}{\big({\textstyle\frac{1}{2}}\beta^*_2\big)^\frac{1}{4}}u\Big\}\Big],~~~
\label{eq:zer*2+}
\end{eqnarray}
where, given the magnitude of $\beta_2$ given by (\ref{eq:kbet-J2}), we  extended the integration limits $\pm\pi \big({\textstyle\frac{1}{2}}\beta^*_2\big)^\frac{1}{4} \rightarrow \pm\infty$.

\begin{figure}
\includegraphics{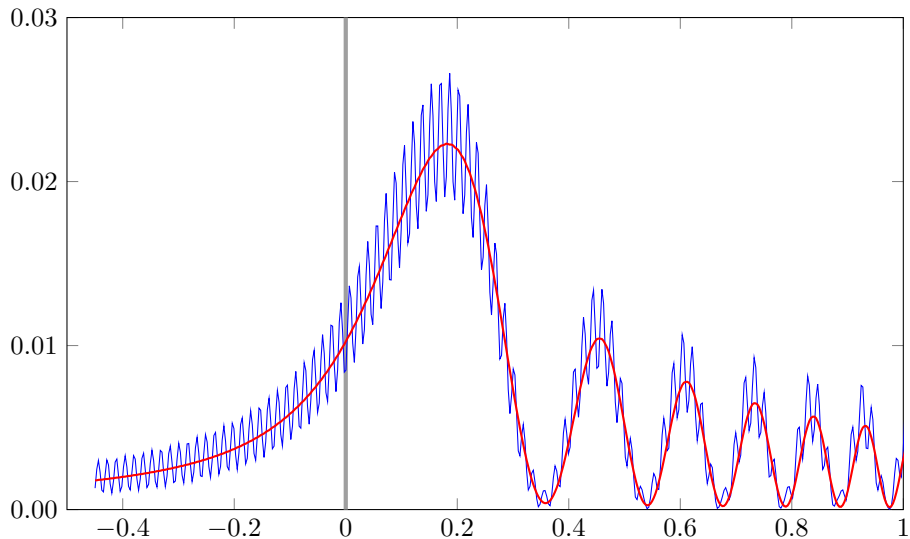}
\caption{\label{fig:caustics-cusp} Behavior of the PSF of the SGL in the vicinity of the caustic cusp, at $\phi=0$, as modeled by (\ref{eq:zer*2+}). The caustic boundary is marked by a thick gray vertical bar at the origin of the coordinate system. The parameterization corresponds to that of Fig.~\ref{fig:caustics-cusp-all} left, but a shorter wavelength (250~nm) is used to better illustrate the high spatial frequency oscillations modulating the caustic pattern. Note that the optical axis is to the right of this image, $\approx 2.55$~m from the origin.
}
\end{figure}

In this expression, we recognize the Pearcey integral (\ref{eq:Pearcey}), which is well-studied in the context of the cuspoid catastrophe  \cite{Pearcey:1946,Connor-Farrelly:1981,Berry-Howles:2010}. In terms of the Pearcey integral, the complex amplitude of the EM field takes the form
{}
\begin{eqnarray}
B^*_2(\overline \rho, \phi)&=&\frac{1}{\big({\textstyle\frac{1}{2}}\beta_2+\sum_{n=2}^\infty {\textstyle\frac{2}{3}}n^3\beta_{2n}\big)^\frac{1}{4}}\frac{e^{i \varphi_0}}{{2\pi}}
\Pe(\overline x,\overline y),
\label{eq:perB}
\end{eqnarray}
where  $\overline x$ and $\overline y$ are given by
\begin{eqnarray}
\overline x =-
\frac{{\textstyle\frac{1}{2}}\alpha\widetilde\rho +3\beta_2\phi^2+\sum_{n=2}^\infty 2 n^2\beta_{2n}}{\big({\textstyle\frac{1}{2}}\beta_2+\sum_{n=2}^\infty {\textstyle\frac{2}{3}}n^4\beta_{2n}\big)^\frac{1}{2}}, \qquad
\overline y =
\frac{(\alpha\widetilde\rho -4\beta_2\phi^2)\phi}{\big({\textstyle\frac{1}{2}}\beta_2+\sum_{n=2}^\infty {\textstyle\frac{2}{3}}n^4\beta_{2n}\big)^\frac{1}{4}}.
\label{eq:perPSF}
\end{eqnarray}

As a result, the PSF in the direction of the $J_2$ cusp is given as
{}
\begin{eqnarray}
{\tt PSF}(\widetilde\rho,\phi)= B^*_2(\widetilde\rho,\phi)B_2(\widetilde\rho,\phi)&=&\frac{1}{\big({\textstyle\frac{1}{2}}\beta_2+\sum_{n=2}^\infty {\textstyle\frac{2}{3}}n^4\beta_{2n}\big)^\frac{1}{2}}\frac{1}{(2\pi)^2}\big| \Pe(\overline x,\overline y)\big|^2.
\label{eq:per}
\end{eqnarray}

Fig.~\ref{fig:caustics-cusp} shows the comparison between the PSF in the direction of the cusp, $\phi=0$, as given by the complete diffraction integral (\ref{eq:ph*0}) (or, effectively, (\ref{eq:B2c}) with $\phi=0$ and $\phi_s$ absent) and the PSF modeled by (\ref{eq:zer*2+}). As we can see, the Pearcey integral appears to correctly model the averaged PSF not only at the caustic boundary but also inside and outside of it. The high-frequency behavior of the PSF (with the approximate spatial frequency of the monopole PSF) is averaged out; the lower-frequency behavior that emerges due to the quadrupole remains. The Pearcey integral fails only near the central region, where the complete diffraction integral shows elevated light levels due to the residual effects of the monopole PSF, not completely canceled out by the quadrupole; this elevation is not captured by the Pearcey integral. Despite this good agreement between the complete diffraction integral and the Pearcey integral, we need to offer a word of caution: this good agreement exists only in the direction of the cusp, $\phi=0$.

The square of the Pearcey integral reaches its maximum of $\big| \Pe(\overline x,0)\big|^2\simeq6.94$ at the point $\overline x=-2.19,\overline y=0$. Thus, if only $J_2$ is considered, the largest value of the PSF estimated at that point is
{}
\begin{eqnarray}
{\tt PSF}_{\tt cusp}
\lesssim \sqrt{\frac{2}{\pi \beta_2}}\,\frac{1}{\pi},
\label{eq:per=}
\end{eqnarray}
which is occurring at the position that is by $\widetilde\rho=2.19\,({\textstyle\frac{1}{2}}\beta_2)^\frac{1}{2}/{\textstyle\frac{1}{2}}\alpha=3.75~{\rm m}\, ({\lambda}/{1\,\mu{\rm m}})^\frac{1}{2}|\sin\beta_s|$ closer to the optical axis as measured from the caustic boundary at the cusp, at $\rho_2=4\beta_2/\alpha$; cf. Eq.~(\ref{eq:mag-J2}).

We note that the caustic does not correspond to the position of the maximum intensity of the PSF on the image plane at a particular heliocentric distance. The Pearcey integral approximation, shown in Fig.~\ref{fig:caustics-cusp}, compared against the position of the caustic boundary, shows that the boundary corresponds to the integral's last inflection point. This position appears to mark the transition from a region dominated by the caustic pattern to a region dominated by the concentric Airy pattern characteristic of the monopole PSF.

Considering the presence of other zonal harmonics shown in (\ref{eq:perPSF}) and (\ref{eq:per}), and treating $\beta_s=1$ (i.e., the source is at the solar equator -- the most conservative case), we find that the maximum of the {\tt PSF} increases by a factor of $\sim 1.22$. The shape of the peak at the cusp widens and its maximum moves closer  to the optical axis by $\sim 8.36$~m, driven by the contribution from the octupole, $J_4$. However, for  $\beta_s=0.1$, the change in the  {\tt PSF} peak is negligible; it shifts towards the optical axis only by $\sim 0.36$~m. It appears that although $J_2$ produces the most pronounced effect on the position and the shape of the cusp, the contribution from the octupole, $J_4$, must also be considered for high-resolution image reconstruction. Contributions of other solar zonal harmonics are negligible.

\subsection{PSF in the vicinity of  the fold in the valley between the cusps}
\label{sec:wo-fold}

Similarly to the discussion in Section~\ref{sec:wo-cusp},  we begin with  the PSF in the vicinity of the fold of the quadrupole caustic. By setting in (\ref{eq:ph*f1})  $\phi={\textstyle\frac{\pi}{4}}$, we obtain the expression for the phase of a complex conjugate of the amplitude $B$ from (\ref{eq:zer*1}) as
{}
\begin{eqnarray}
\varphi_2(\overline \phi_\xi,{\textstyle\frac{\pi}{4}})&=&-2\beta_2\cos\overline\phi_\xi\Big(\sin\overline\phi_\xi -\Big(1-\frac{\alpha\widetilde\rho}{2\beta_2}\Big)\Big).
\label{eq:ph*0p4}
\end{eqnarray}

Next, we consider the stationary points of the $J_2$ caustic. The stationary points are given by computing the first derivative  $d\varphi_2(\overline \phi_\xi,{\textstyle\frac{\pi}{4}})/d{\phi_\xi}$ that is done as
{}
\begin{eqnarray}
\frac{d\varphi_2(\overline \phi_\xi,{\textstyle\frac{\pi}{4}})}{d\phi_\xi}&=&4\beta_2\Big\{\Big(\sin\overline\phi_\xi -{\textstyle\frac{1}{4}}\Big(1-\frac{\alpha\widetilde\rho}{2\beta_2}\Big)\Big)^2- {\textstyle\frac{1}{2}}-\Big({\textstyle\frac{1}{4}}\Big(1-\frac{\alpha\widetilde\rho}{2\beta_2}\Big)\Big)^2\Big\}.
\label{eq:ph-der*0p4}
\end{eqnarray}

Equating this expression to zero, we determine the stationary points and those with highly-oscillating behavior, indicating the presence of the fold. Thus, at the fold, where $\widetilde\rho=0$,  equation $d\varphi_2(\overline \phi_\xi,{\textstyle\frac{\pi}{4}})/d{\phi_\xi}=0$ yields two equations to determine $\overline \phi_\xi$, namely  $\sin\overline \phi_\xi=1$ and $\sin\overline \phi_\xi=-{\textstyle\frac{1}{2}}$, yielding $\overline \phi_\xi={\textstyle\frac{\pi}{2}}$ and $\overline \phi_\xi=-{\textstyle\frac{\pi}{6}}$. To identify the appropriate solution that indicates the presence of the fold, we need to study the behavior of the second derivative of the phase $d^2\varphi_2(\overline \phi_\xi,{\textstyle\frac{\pi}{4}})/d{\phi_\xi^2}$, that is computed from (\ref{eq:ph-der*0p4}) as
{}
\begin{eqnarray}
\frac{d^2\varphi_2(\overline \phi_\xi,{\textstyle\frac{\pi}{4}})}{d\phi_\xi}&=&8\beta_2\cos\overline\phi_\xi\Big(\sin\overline\phi_\xi -{\textstyle\frac{1}{4}}\Big(1-\frac{\alpha\widetilde\rho}{2\beta_2}\Big)\Big).
\label{eq:ph-der*03p4}
\end{eqnarray}

Considering (\ref{eq:ph-der*03p4}) at the fold, $\widetilde\rho=0$, we substitute the solutions $\overline \phi_\xi={\textstyle\frac{\pi}{2}}$ and $\overline \phi_\xi=-{\textstyle\frac{\pi}{6}}$ and consider the behavior of  the second derivative. We see that $\overline \phi_\xi=-{\textstyle\frac{\pi}{6}}$  results in a regular behavior of the second derivative. However, the solution $\overline \phi_\xi={\textstyle\frac{\pi}{2}}$  causes the second derivative to vanish at the fold, $\widetilde\rho=0$. Thus, in the vicinity of the fold, most of the contribution to the phase  comes from the area where $\overline\phi_\xi \approx {\textstyle\frac{\pi}{2}}$. This observation allows us to consider behavior of the phase (\ref{eq:ph*f1}) around the point $\overline\phi_\xi ={\textstyle\frac{\pi}{2}}$. To proceed, we introduce a new variable $\widetilde\phi_\xi$ in accord to
{}
\begin{eqnarray}
\overline\phi_\xi ={\textstyle\frac{\pi}{2}}+\widetilde\phi_\xi.
\label{eq:w-tilde-p4}
\end{eqnarray}

Similarly to the approach we took to derive (\ref{eq:ph*df}), we will use the complete phase at the caustic given by (\ref{eq:ph*f1}) and develop the needed approximation. For that, we substitute (\ref{eq:w-tilde-p4}) in (\ref{eq:ph*f1}) and expand the result in the power series of the small angle  $\widetilde\phi_\xi$, retaining only the leading terms, which yields
{}
\begin{eqnarray}
\varphi(\widetilde \phi_\xi,\phi)&=&-\beta_2\Big(\cos 2\phi+\sin 4\phi\Big)+ \Big(\alpha\widetilde\rho +\beta_2\big(2\sin2\phi-\cos4\phi -3\big)\Big)\widetilde\phi_\xi+\nonumber\\
&+&{\textstyle\frac{1}{2}}\beta_2\Big(4\cos2\phi +\sin 4\phi\Big)\widetilde\phi^2_\xi+
 {\textstyle\frac{1}{6}}\Big(-\alpha\widetilde\rho +\beta_2 \big(3+\cos4\phi-8\sin2\phi\big)\Big)\widetilde\phi^3_\xi+{\cal O}(\widetilde\phi^4_\xi),
\label{eq:ph*dfp4}
\end{eqnarray}
where we do not account for the presence of the multipolar terms, as their contribution at the fold is negligible.

Next, we are interested to study the behavior of the complex amplitude in the small vicinity of the central fold that is around $\phi={\textstyle\frac{\pi}{4}}$. To do that, we rotate the coordinates by introducing the small angle $\overline\phi\ll1$
{}
\begin{eqnarray}
\phi ={\textstyle\frac{\pi}{4}}+\overline\phi,
\label{eq:w-phi}
\end{eqnarray}
transforming (\ref{eq:ph*dfp4}) while expanding the result in terms of the small parameter $\overline\phi$:
{}
\begin{eqnarray}
\varphi(\widetilde \phi_\xi,\overline\phi)&=&-(\beta_2+\tfrac{1}{6}\alpha\widetilde\rho)\cdot\widetilde\phi^3_\xi-6\beta_2\overline\phi\cdot\widetilde\phi^2_\xi+ \alpha\widetilde\rho\widetilde\phi_\xi+6\beta_2\overline\phi+{\cal O}(\widetilde\phi^4_\xi).
\label{eq:ph*75d}
\end{eqnarray}

Using the definitions of $\overline\phi_\xi$, $\widetilde\phi_\xi$ and $\overline\phi$, given by (\ref{eq:phxibar}), (\ref{eq:w-tilde-p4}), and (\ref{eq:w-phi}) correspondingly, we have $\widetilde\phi_\xi=\phi_\xi-{\textstyle\frac{3\pi}{4}}-\overline\phi$. Recognizing this relationship between the angles and remembering that the angle $\overline\phi$ is small, we transform (\ref{eq:ph*75d}) as
{}
\begin{eqnarray}
\varphi( \phi_\xi,\overline\phi)&=& -(\beta_2+\tfrac{1}{6}\alpha\widetilde\rho)\,\big(\phi_\xi-{\textstyle\frac{3\pi}{4}}\big)^3 -(3\beta_2-\tfrac{1}{2}\alpha\widetilde\rho)\overline\phi\,\big(\phi_\xi-{\textstyle\frac{3\pi}{4}}\big)^2+ \Big(\alpha\widetilde\rho -3\beta_2 \overline\phi^2\Big)\big(\phi_\xi-{\textstyle\frac{3\pi}{4}}\big)+w_1+{\cal O}(\widetilde\phi^4_\xi),
\label{eq:ph*ff}
\end{eqnarray}
where the constant $w_1$ is given as $w_1=(6\beta_2-\alpha\widetilde\rho)\overline\phi$.

We note that expression (\ref{eq:ph*ff}) bears close resemblance to the phase of the Airy integral (\ref{eq:Airy}). To present it in that form, we need to depress the cubic structure in (\ref{eq:ph*ff}). This can be done by introducing yet another transformation of variables $\phi_\xi-{\textstyle\frac{3\pi}{4}}=\widehat\phi_\xi-[(\beta_2-\tfrac{1}{6}\alpha\widetilde\rho)/(\beta_2+\tfrac{1}{6}\alpha\widetilde\rho)]\overline\phi$, which yields
{}
\begin{eqnarray}
\varphi(\widehat \phi_\xi,\phi)&=& -(\beta_2+\tfrac{1}{6}\alpha\widetilde\rho)\,\widehat\phi_\xi^3 + \alpha\widetilde\rho \, \widehat\phi_\xi+\varphi_2+{\cal O}(\widetilde\phi^4_\xi),
\label{eq:ph*fppf}
\end{eqnarray}
where the constant $\varphi_2$ is given as $\varphi_2=6\beta_2[(\beta_2-\tfrac{1}{6}\alpha\widetilde\rho)/(\beta_2+\tfrac{1}{6}\alpha\widetilde\rho)]\overline\phi$.

Finally, we introduce the variable $u$, needed to integrate (\ref{eq:zer*1}):
{}
\begin{align}
(\beta_2+\tfrac{1}{6}\alpha\widetilde\rho)\widehat\phi_\xi^3\, =\, \tfrac{1}{3}u^3 \qquad \Rightarrow \qquad
u&\, =\, \Big\{\phi_\xi-{\textstyle\frac{3\pi}{4}}+\Big((\beta_2-\tfrac{1}{6}\alpha\widetilde\rho)/(\beta_2+\tfrac{1}{6}\alpha\widetilde\rho)\Big)\overline\phi\Big\}\big(
3\beta_2+\tfrac{1}{2}\alpha\widetilde\rho
\big)^\frac{1}{3}, \qquad
\nonumber\\
\phi_\xi&\, =\, {\textstyle\frac{3\pi}{4}}-\Big((\beta_2-\tfrac{1}{6}\alpha\widetilde\rho)/(\beta_2+\tfrac{1}{6}\alpha\widetilde\rho)\Big)\overline\phi+u\big(
3\beta_2+\tfrac{1}{2}\alpha\widetilde\rho
\big)^{-{\frac{1}{3}}}.
\label{eq:var-vddar}
\end{align}

These notations and definitions allow us to transform the phase (\ref{eq:ph*fppf}) as
{}
\begin{eqnarray}
\varphi(u,0)&=&-\tfrac{1}{3}u^3 +\frac{\alpha\widetilde\rho}{(3\beta_2+\tfrac{1}{2}\alpha\widetilde\rho)^\frac{1}{3}}u +\varphi_2.
\label{eq:ph*dd}
\end{eqnarray}

We may now substitute these results into the integral (\ref{eq:zer*1})  that describes the complex conjugate amplitude of the EM field. As a result, in the vicinity of the fold of the quadrupole caustic this amplitude  takes the form:
{}
\begin{eqnarray}
B_2(\widetilde\rho, \phi)&=&\frac{1}{(3\beta_2+\tfrac{1}{2}\alpha\widetilde\rho)^{{\frac{1}{3}}}} \frac{e^{i\varphi_2}}{2\pi} \int_{-\infty}^{\infty} du \exp\Big[i\Big\{\tfrac{1}{3}u^3 -(3\beta_2+\tfrac{1}{2}\alpha\widetilde\rho)^{-\frac{1}{3}}{\alpha\widetilde\rho}\,u\Big\}\Big],~~~
\label{eq:zer*2+fold}
\end{eqnarray}
where we recognize that, given the magnitude of $\beta_2$ given by (\ref{eq:kbet-J2}), we can extend the integration limits $-{\textstyle\frac{3\pi}{4}}(3\beta_2+\tfrac{1}{2}\alpha\widetilde\rho)^\frac{1}{3} \rightarrow -\infty$ and ${\textstyle\frac{5\pi}{4}}(3\beta_2+\tfrac{1}{2}\alpha\widetilde\rho)^\frac{1}{3} \rightarrow +\infty$ without introducing a significant additional error.

This integral is the well-known Airy integral $\Ai(x)$ (\ref{eq:Airy}), with
\begin{eqnarray}
x =-(3\beta_2+\tfrac{1}{2}\alpha\widetilde\rho)^{-\frac{1}{3}}\alpha \widetilde\rho.
\label{eq:perPSFx}
\end{eqnarray}

As a result, when $\beta_2\gg|\alpha\widetilde\rho|$, i.e., in the vicinity of the caustic boundary, the PSF in the direction of the fold of the astroid (i.e, $J_2$) caustic can be given as
{}
\begin{eqnarray}
{\tt PSF}(\widetilde\rho,\phi)= B^*_2(\widetilde\rho,\phi)B_2(\widetilde\rho,\phi)&=&(3\beta_2)^{-\frac{2}{3}}\Ai\big((-3\beta_2)^{-\frac{1}{3}}\alpha\widetilde\rho\big)^2.
\label{eq:perfoldp}
\end{eqnarray}

\begin{figure}
\includegraphics{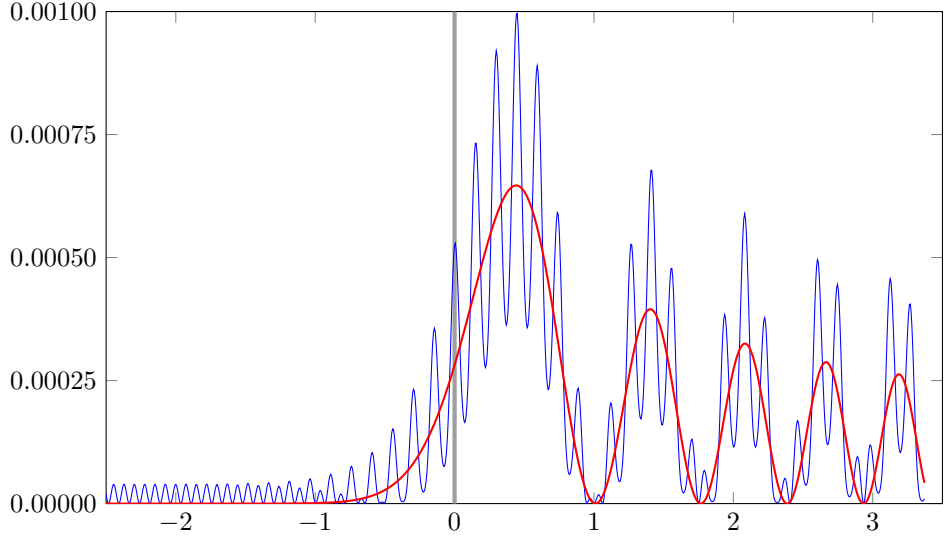}
\caption{\label{fig:caustics-fold-ai} Behavior of the PSF of the quadrupole SGL. The actual PSF, shown in blue, is as prescribed by (\ref{eq:ph*0p4}) (or, effectively, from (\ref{eq:B2c}) with $\phi={\textstyle\frac{\pi}{4}}$ and $\phi_s$ absent.) The PSF, as modeled by (\ref{eq:zer*2+fold}), is shown in red. To highlight the relationship between the approximation and the high spatial frequency oscillations of the actual PSF, a different parameterization is used: $J_2\sin^2\beta_s=2\times 10^{-7}$, with $\lambda=1~\mu$m. Note that the optical axis is to the right of this image, $\approx 127.4$~m from the origin. As in Fig.~\ref{fig:caustics-cusp}, the caustic boundary is marked by a thick gray vertical bar at the origin of the coordinate system.
}
\end{figure}

In the immediate neighborhood of the caustic fold at  $\phi={\textstyle\frac{\pi}{4}}$, the result is independent of the angle $\phi$. It is driven by the simple combinations of the parameters (\ref{eq:perPSFx}). While our model reproduces the overall behavior of the integral  (\ref{eq:B2c}), the high frequency content is missing. Similar to the Pearcey integral characterizing the behavior in the direction of the cusps, we note that the high spatial frequency contribution (characteristic of the monopole) is effectively averaged in the vicinity of the caustic by this approximation.

The square of the Airy function reaches its maximum of $\Ai(x)^2\simeq 0.287$ at $x=-1.02$. The maximum of the PSF estimated at that point is
{}
\begin{eqnarray}
{\tt PSF}_{\tt fold}
\lesssim 0.287\big(3\beta_2\big)^{-\frac{2}{3}}.
\label{eq:air}
\end{eqnarray}

For the SGL, this maximum occurs at the distance of $\rho=1.02\,\big(3\beta_2\big)^{\frac{1}{3}}/\alpha=0.44~{\rm m}\, ({\lambda}/{1\,\mu{\rm m}})^\frac{2}{3}({r}/{650\,{\rm AU}})^\frac{1}{6}|\sin^\frac{2}{3}\beta_s|$ closer to the optical axis as measured from the caustic boundary at the fold, at $\rho_2=2\beta_2/\alpha$; cf. Eq.~(\ref{eq:mag-J2}).

Thus, once again, we see that the peak intensity of the astroid caustic is not exactly at the fold of the caustic, but at a small distance, $\rho\sim{\textstyle\frac{3}{2}}\,\beta_2^{\frac{1}{3}}/\alpha$, towards the optical axis. In general, we observe that the caustic boundary corresponds to the position of the last inflection point in the PSF with its high spatial frequency component averaged. This indeed marks the point of transition from a PSF dominated by the caustic to a PSF dominated by the concentric Airy-pattern of the monopole.

\subsection{Properties of the caustic boundary}

The caustic boundary separates two fundamentally distinct regions of the PSF. The exterior of the boundary is dominated by the concentric pattern that we recognize as being characteristic of the monopole PSF. In contrast, the interior is characterized by a pattern that is determined by the multipole moments.

We recognize two distinct spatial wavelengths. The short wavelength that is representative of the monopole pattern far outside the caustic boundary also survives in the interior, but only as a pattern that modulates a dominant, longer-wavelength oscillation that characterizes the interior region.

As we observe looking at Figs.~\ref{fig:caustics-cusp} and \ref{fig:caustics-fold-ai}, the caustic boundary is located at the last inflection point of this longer-wavelength oscillation. Inside this region the oscillatory behavior of the PSF is dominated by the multipole moments; outside this boundary, the pattern rapidly settles down to that of the monopole PSF.

\begin{figure}
\includegraphics{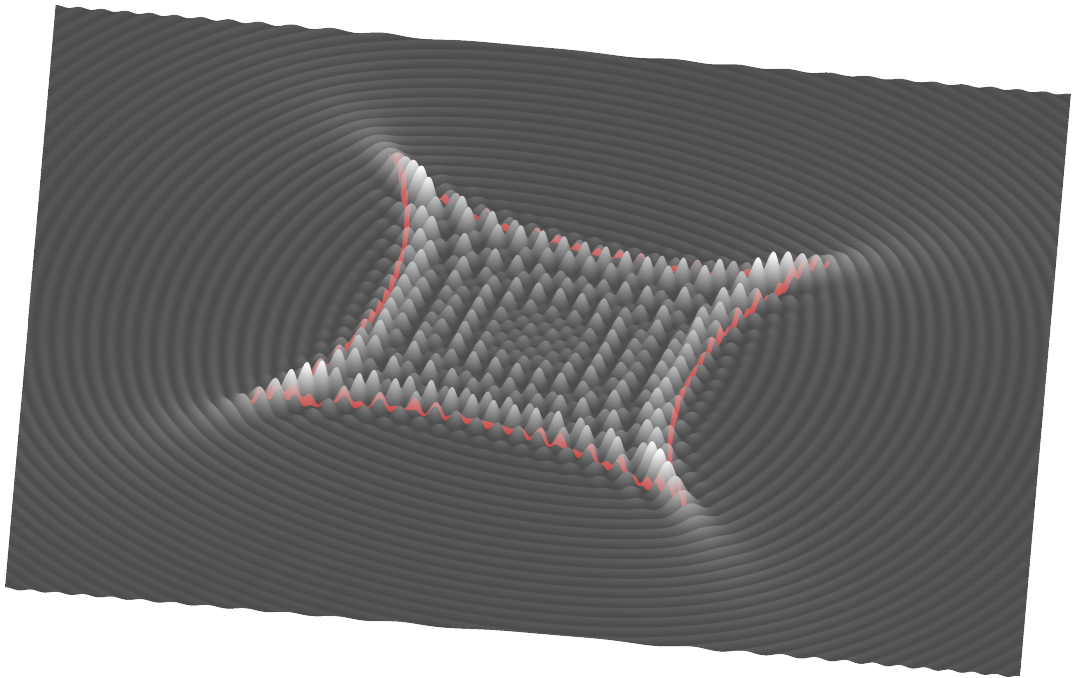}
\caption{\label{fig:persp-caustic}Another view of the astroid caustic boundary (red), projected onto a surface plot of the PSF, which was generated using the same parameterization as Fig.~\ref{fig:caustics-cusp-all}. We note that the magnitude of the PSF varies strongly along the caustic boundary.}
\end{figure}

The caustic boundary itself is characterized by a rapidly oscillating amplitude. This is visually demonstrated in Fig.~\ref{fig:persp-caustic}, in which the astroid caustic boundary is superimposed on a three-dimensional representation of the magnitude of the PSF. We easily recognize the concentric pattern, familiar from the monopole PSF, rapidly emerging outside the caustic boundary. Just inside the boundary, however, we encounter the maxima of the PSF, most pronounced in the cusp regions.

\begin{figure}
\includegraphics{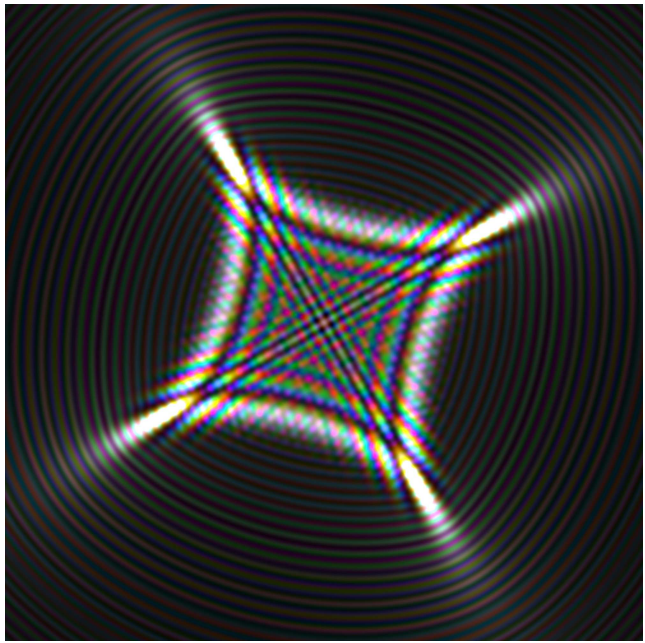}
\caption{\label{fig:rainbow}Wavelength dependence of the astroid caustic, shown using colors of the rainbow. Adapted from \cite{Turyshev-Toth:2021-multipoles}.}
\end{figure}

Finally, it is instructive to look again at Fig.~{\ref{fig:rainbow}} (adapted from \cite{Turyshev-Toth:2021-multipoles}). This color figure demonstrates the extent to which the caustic pattern depends on wavelength. We note that the cusps are dominated by white: the location of the cusps is independent of wavelength, and the magnitude of the signal is much larger than any wavelength-dependent oscillations that modulate it. Elsewhere, in contrast, the colors of the rainbow emerge as the spatial wavelengths of the various oscillating patterns that we see depend on $k$. This remains true outside the caustic boundary as well, since the Airy-pattern associated with the monopole PSF is also wavelength-dependent; this dependence is lost only if we average the signal over an area that is significantly larger than the spatial wavelength, as demonstrated in Ref.~\cite{Turyshev-Toth:2021-multipoles}.

\subsection{Computing the PSF}

In the preceding subsections, we presented a series of approximations that shed light on the behavior of the PSF of an axially symmetric gravitational lens, with special emphasis on the dominating quadrupole moment. We also established relationships with previously known results from the literature.

In these discussions, we were able to make comparisons between approximations and the true value of the PSF because of the power of the angular eikonal integral~(\ref{eq:zer*1}). Though this integral requires numerical evaluation, in the vicinity of the optical axis, this evaluation can be accomplished using a modest number of integration steps. As a result, Eq.~(\ref{eq:zer*1}) can be readily used to compute the PSF in the presence of an arbitrary combination of the $J_2..J_8$ multipoles (higher multipoles could be included just as easily, but for the SGL, they do not contribute significantly).

We implemented Eq.~(\ref{eq:zer*1}) in the form of a program in the {\tt C++} language, calculating the PSF, that is to say, light from a point source, projected by the SGL onto the image plane. The program is parameterized by the location (relative to the optical axis) and size of the area of interest in the image plane and the desired resolution. Physical parameters include the wavelength, the distance $r$ from the Sun, and the direction with respect to the solar axis of rotation, as characterized by $\sin\beta_s$, with the angle $\beta_s$ defined by (\ref{eq:note}). Additionally, the program allows us to manually adjust the values of $J_{2..8}$ individually, in order to explore the magnitude of the relative contributions of these coefficients to image formation. The output of this program is a binary array of floating-point numbers, representing light intensity in the image plane sampled at the requested resolution.

When only even-numbered zonal harmonics are present (which is the case for the Sun), computational time can be saved by calculating only one quadrant of the PSF; the other three quadrants can then be obtained using appropriately rotated copies of the quadrant that was calculated. A utility program that accomplishes this is one of several utilities we developed for the post-processing of images using command-line pipelining. Finally, images are converted from the floating-point representation to standard formats using the open-source software package {\tt ImageMagick}, which has the ability to convert from user-defined formats to standards such as {\tt JPEG} or {\tt PNG}.

To generate the three-dimensional Fig.~\ref{fig:persp-caustic}, additional tools were required, as off-the-shelf plotting software could not readily render the caustic boundary onto the extruded ``3D'' representation of the PSF. For this, we employed standard graphics processing algorithms to rotate, extrude, and project a precomputed representation of the PSF, while superimposing the (also precomputed) caustic boundary onto the resulting image. This utility joins a growing library of short programs that we are constructing to efficiently process various representations of the PSF and the resulting images, for future study. The resulting data sets can also be readily processed using the {\tt ffmpeg} software library of subroutines and utilities to create animated presentations. Additional extensions, which allow our software tools to represent the SGL far from the optical axis, and which allow it to be convolved with the PSF of an optical telescope to study the resulting formation of the (partial or full) Einstein-ring around the Sun are also under development and will be used in our future work.

\section{Discussion and Conclusions}
\label{sec:end}

In this paper we studied the caustics formed by a realistic gravitational lens using the angular eikonal method, with particular attention paid to the case of the solar gravitational lens, the SGL. As we remarked in \cite{Turyshev-Toth:2021-multipoles}, nothing is perfect in our Universe, not even the Sun. Its gravitational field is not perfectly spherically symmetric. In this paper we have studied the  optical properties of the axisymmetric SGL described in terms of zonal harmonics, and the caustics formed by the diffraction of light in solar gravity field. To this end, we developed a wave-optical treatment of light propagation in the vicinity of the Sun.
This work is important for our on-going efforts on studying the SGL as the means for multipixel imaging of exoplanets \cite{Turyshev-Toth:2020-extend,Toth-Turyshev:2020}  in the context of a realistic space mission \cite{Turyshev-etal:2020-PhaseII}.

As we have demonstrated here, oblateness (i.e., nonvanishing zonal harmonics) changes the structure and thus the optical properties of the SGL. Its PSF is now given by a set of superimposed caustics, each with a unique set of cusps and folds. As a result, light from a point source is not focused around a point (as in the case of the monopole with ${\tt PSF}_0(\vec x)=J^2_0(k\sqrt{{2r_g}/{r}}\rho)$), but it is distributed over the caustic region.

The approach presented here may used to quantify the amount of light deposited at various locations in the image plane. Specifically, we can determine
\begin{inparaenum}[i)]
\item how much light is deposited along the caustic (including the cusps and folds),
\item how much light falls in the interior region of the caustic, and
\item how much light is scattered outside the caustic boundary.
\end{inparaenum}
These questions are key to the development of an improved understanding of the optical properties of the SGL of a realistic Sun.  The methods developed in this paper can be applied to study the caustic formed by zonal harmonics. In the case of the Sun, the largest caustic contribution comes from the quadrupole moment $J_2$. This led us to focus our studies on the astroid caustic in particular.

We considered diffraction of light in the gravitational field of the Sun and also interference effects in the image plane. The combination of diffraction and interference leads to the formation of caustics in the image plane of the SGL. Diffraction results in different optical paths taken by light rays as they pass by the Sun, enveloping it on all sides. Once these rays reach the image plane, they interfere either constructively or destructively. Axial symmetry of the solar gravitational field makes it possible to capture this interference process using the formalism of the dimensionless zonal harmonic coefficients $J_n$.

Using this formalism, we were able to describe the caustic patterns projected by an extended gravitational field, such as the Sun's, in an image plane in the strong interference region of the gravitational lens. It has been known that the intensity of light in the inner and outer regions of the caustic can be modeled using the methods of the geometric optics, WKB and stationary phase. Such results can provide a reasonable approximation. However, the EM field in the immediate vicinity of the caustic boundary must be modeled using wave-optical methods.

In this paper, we developed a suitable approximate solution that accurately describes the EM field everywhere in the strong interference region, including the immediate vicinity of the cusps and folds along the caustic boundary. Our result is generic and can describe any combination of axisymmetric multipoles (i.e., any gravitational field described using zonal harmonics\footnote{Though the derivation is tedious, the approach can also be extended to describe lensing by arbitrary weak monopole-dominated gravitational fields, represented using zonal and tesseral harmonics, i.e., unconstrained spherical harmonic coefficients.}). We also demonstrated that for the special case of the quadrupole moment, the astroid caustic, our results easily replicate the well-known Pearcy and Airy integrals, which describe the approximate behavior of the gravitational lens near the projected cusps and folds, from the diffraction catastrophe theory.

A similar approach may be used to study any $J_n$ caustic, as described by the diffraction integral (\ref{eq:B2}). The Pearcey (cusp) and Airy (fold) diffraction catastrophes are just the first members of a hierarchy of wave patterns decorating the caustic singularities classified by catastrophe theory. The diffraction catastrophes are complicated, but useful information is contained in scaling laws describing how the intensity increases and the fringe spacings decrease as the wavelength $\lambda$ gets smaller. Considering (\ref{eq:zer*2+}), we see that for the cusp, the intensity increases as $\lambda^{-\frac{1}{2}}$ and the fringes shrink as $\lambda^{\frac{1}{2}}$ along the cusp and $\lambda^{\frac{3}{4}}$ across it. For the fold, we examine (\ref{eq:zer*2+fold}), to see that the intensity increases as $\lambda^{-\frac{1}{3}}$ and the fringes shrink as $\lambda^{\frac{2}{3}}$ across it. This information will be useful for the study of imaging with the SGL.

The caustics that these processes form are aesthetically pleasing because they represent the natural symmetries of an axisymmetric gravitational field. Not only do the patterns that form in the image plane possess information on the optical properties of the lens, the opposite is also true -- this information may be used to reconstruct the lens. 
This aspect of our approach may be of critical importance for many areas of modern astrophysics, especially those relying on gravitational microlensing to study the structure and the composition of distant lensing objects in the Universe.  

The results presented here may be used to describe gravitational lensing by realistic astrophysical lenses, including stars, spiral and elliptical galaxies. Description of lensing with an elliptical mass distribution generally faces challenges in evaluating the deflection angles and magnification matrices, ultimately requiring numerical efforts  \cite{Barkana:1998}.  However, if the external gravitational potential of an object may be given in the form of an infinite set of zonal harmonics, the complex amplitude of the resulting EM field is (\ref{eq:B2}). Even that model may further be generalized to describe a generic matter distribution with exterior gravitational potential possessing a more complex structure that may be captured by an infinite set of symmetric trace-free (STF) multipole moments \cite{Turyshev-Toth:2021-multipoles}.  As a result, with the approach presented here, the challenges above become manageable in a semi-analytical manner within a complete wave-optical treatment. 
This applicability of our approach for general astrophysics is important and it is currently being investigated. 

Concerning imaging with the SGL, our results presented here can be used to evaluate the amount of light that is received in the image plane from extended objects such as an exoplanet, which is the subject of our ongoing work. Results, when available, will be reported elsewhere.

\begin{acknowledgments}
This work in part was performed at the Jet Propulsion Laboratory, California Institute of Technology, under a contract with the National Aeronautics and Space Administration. VTT acknowledges the generous support of Plamen Vasilev and other Patreon patrons.

\end{acknowledgments}


\begin{thebibliography}{32}
\expandafter\ifx\csname natexlab\endcsname\relax\def\natexlab#1{#1}\fi
\expandafter\ifx\csname bibnamefont\endcsname\relax
  \def\bibnamefont#1{#1}\fi
\expandafter\ifx\csname bibfnamefont\endcsname\relax
  \def\bibfnamefont#1{#1}\fi
\expandafter\ifx\csname citenamefont\endcsname\relax
  \def\citenamefont#1{#1}\fi
\expandafter\ifx\csname url\endcsname\relax
  \def\url#1{\texttt{#1}}\fi
\expandafter\ifx\csname urlprefix\endcsname\relax\def\urlprefix{URL }\fi
\providecommand{\bibinfo}[2]{#2}
\providecommand{\eprint}[2][]{\url{#2}}

\bibitem[{\citenamefont{{Turyshev} and {Toth}}(2017)}]{Turyshev-Toth:2017}
\bibinfo{author}{\bibfnamefont{S.~G.} \bibnamefont{{Turyshev}}}
  \bibnamefont{and} \bibinfo{author}{\bibfnamefont{V.~T.}
  \bibnamefont{{Toth}}}, \bibinfo{journal}{Phys. Rev. D}
  \textbf{\bibinfo{volume}{96}}, \bibinfo{pages}{024008}
  (\bibinfo{year}{2017}), \eprint{arXiv:1704.06824 [gr-qc]}.

\bibitem[{\citenamefont{Liebes}(1964)}]{Liebes:1964}
\bibinfo{author}{\bibfnamefont{S.}~\bibnamefont{Liebes}},
  \bibinfo{journal}{Phys. Rev.} \textbf{\bibinfo{volume}{133}},
  \bibinfo{pages}{B835} (\bibinfo{year}{1964}).

\bibitem[{\citenamefont{{Deguchi} and {Watson}}(1986)}]{Deguchi-Watson:1986}
\bibinfo{author}{\bibfnamefont{S.}~\bibnamefont{{Deguchi}}} \bibnamefont{and}
  \bibinfo{author}{\bibfnamefont{W.~D.} \bibnamefont{{Watson}}},
  \bibinfo{journal}{Astrophys. J.} \textbf{\bibinfo{volume}{307}},
  \bibinfo{pages}{30} (\bibinfo{year}{1986}).

\bibitem[{\citenamefont{{Schneider} et~al.}(1992)\citenamefont{{Schneider},
  {Ehlers}, and {Falco}}}]{Schneider-Ehlers-Falco:1992}
\bibinfo{author}{\bibfnamefont{P.~S.} \bibnamefont{{Schneider}}},
  \bibinfo{author}{\bibfnamefont{J.}~\bibnamefont{{Ehlers}}}, \bibnamefont{and}
  \bibinfo{author}{\bibfnamefont{E.}~\bibnamefont{{Falco}}},
  \emph{\bibinfo{title}{Gravitational Lenses}}
  (\bibinfo{publisher}{Springer-Verlag Berlin Heidelberg},
  \bibinfo{year}{1992}).

\bibitem[{\citenamefont{{Refsdal} and {Surdej}}(1994)}]{Refsdal-Surdej:1994}
\bibinfo{author}{\bibfnamefont{S.}~\bibnamefont{{Refsdal}}} \bibnamefont{and}
  \bibinfo{author}{\bibfnamefont{J.}~\bibnamefont{{Surdej}}},
  \bibinfo{journal}{Reports on Progress in Physics}
  \textbf{\bibinfo{volume}{57}}, \bibinfo{pages}{117} (\bibinfo{year}{1994}).

\bibitem[{\citenamefont{{Narayan} and
  {Bartelmann}}(1996)}]{Narayan-Bartelmann:1996}
\bibinfo{author}{\bibfnamefont{R.}~\bibnamefont{{Narayan}}} \bibnamefont{and}
  \bibinfo{author}{\bibfnamefont{M.}~\bibnamefont{{Bartelmann}}}, in
  \emph{\bibinfo{booktitle}{13th Jerusalem Winter School in Theoretical
  Physics: Formation of Structure in the Universe Jerusalem, Israel, 27
  December 1995 - 5 January 1996}} (\bibinfo{year}{1996}),
  \eprint{astro-ph/9606001}.

\bibitem[{\citenamefont{{Kovner}}(1987)}]{Kovner:1987}
\bibinfo{author}{\bibfnamefont{I.}~\bibnamefont{{Kovner}}},
  \bibinfo{journal}{Astrophys. J.} \textbf{\bibinfo{volume}{312}},
  \bibinfo{pages}{22} (\bibinfo{year}{1987}).

\bibitem[{\citenamefont{Chu et~al.}(2016)\citenamefont{Chu, Li, Lin, and
  Pan}}]{Chu:2016}
\bibinfo{author}{\bibfnamefont{Z.}~\bibnamefont{Chu}},
  \bibinfo{author}{\bibfnamefont{G.~L.} \bibnamefont{Li}},
  \bibinfo{author}{\bibfnamefont{W.~P.} \bibnamefont{Lin}}, \bibnamefont{and}
  \bibinfo{author}{\bibfnamefont{H.~X.} \bibnamefont{Pan}},
  \bibinfo{journal}{MNRAS} \textbf{\bibinfo{volume}{461}},
  \bibinfo{pages}{4466} (\bibinfo{year}{2016}).

\bibitem[{\citenamefont{{Blandford} and
  {Narayan}}(1986)}]{Blandford-Narayan:1986}
\bibinfo{author}{\bibfnamefont{R.~D.} \bibnamefont{{Blandford}}}
  \bibnamefont{and}
  \bibinfo{author}{\bibfnamefont{R.}~\bibnamefont{{Narayan}}},
  \bibinfo{journal}{ApJ.} \textbf{\bibinfo{volume}{310}}, \bibinfo{pages}{568}
  (\bibinfo{year}{1986}).

\bibitem[{\citenamefont{{Blandford} and {Kovner}}(1988)}]{Blandford-Kovner1988}
\bibinfo{author}{\bibfnamefont{R.~D.} \bibnamefont{{Blandford}}}
  \bibnamefont{and} \bibinfo{author}{\bibfnamefont{I.}~\bibnamefont{{Kovner}}},
  \bibinfo{journal}{Phys. Rev. A} \textbf{\bibinfo{volume}{38}},
  \bibinfo{pages}{4028} (\bibinfo{year}{1988}).

\bibitem[{\citenamefont{An}(2005)}]{An:2005}
\bibinfo{author}{\bibfnamefont{J.~H.} \bibnamefont{An}},
  \bibinfo{journal}{MNRAS} \textbf{\bibinfo{volume}{356}},
  \bibinfo{pages}{1409} (\bibinfo{year}{2005}).

\bibitem[{\citenamefont{{Schneider} et~al.}(2006)\citenamefont{{Schneider},
  {Kochanek}, and {Wambsganss}}}]{Schneider-etal:2004}
\bibinfo{author}{\bibfnamefont{P.}~\bibnamefont{{Schneider}}},
  \bibinfo{author}{\bibfnamefont{C.}~\bibnamefont{{Kochanek}}},
  \bibnamefont{and}
  \bibinfo{author}{\bibfnamefont{J.}~\bibnamefont{{Wambsganss}}},
  \emph{\bibinfo{title}{Gravitational Lensing:Strong,Weak and Micro}}
  (\bibinfo{publisher}{Springer-Verlag: Berlin}, \bibinfo{year}{2006}).

\bibitem[{\citenamefont{{Keller}}(1995)}]{Keller:1995}
\bibinfo{author}{\bibfnamefont{J.}~\bibnamefont{{Keller}}}, in
  \emph{\bibinfo{booktitle}{Proc. of the Int. Congress of Mathematicians,
  Z\"urich, Switzerland 1994}}, edited by
  \bibinfo{editor}{\bibfnamefont{S.}~\bibnamefont{{Chatterji}}}
  (\bibinfo{publisher}{Birkh\"auser Verlag}, \bibinfo{address}{Basel,
  Switzerland}, \bibinfo{year}{1995}), pp. \bibinfo{pages}{106--119}.

\bibitem[{\citenamefont{{Berry} and {Upstill}}(1982)}]{Berry-Upstill:1982}
\bibinfo{author}{\bibfnamefont{M.~V.} \bibnamefont{{Berry}}} \bibnamefont{and}
  \bibinfo{author}{\bibfnamefont{C.}~\bibnamefont{{Upstill}}},
  \bibinfo{journal}{Optics Laser Technology} \textbf{\bibinfo{volume}{14}},
  \bibinfo{pages}{257} (\bibinfo{year}{1982}).

\bibitem[{\citenamefont{{Turyshev} and
  {Toth}}(2021)}]{Turyshev-Toth:2021-multipoles}
\bibinfo{author}{\bibfnamefont{S.~G.} \bibnamefont{{Turyshev}}}
  \bibnamefont{and} \bibinfo{author}{\bibfnamefont{V.~T.}
  \bibnamefont{{Toth}}}, \bibinfo{journal}{Phys. Rev. D}
  (\bibinfo{year}{2021}), \bibinfo{note}{arXiv:2102.03891 [gr-qc]}.

\bibitem[{\citenamefont{Turyshev and Toth}(2019)}]{Turyshev-Toth:2019}
\bibinfo{author}{\bibfnamefont{S.~G.} \bibnamefont{Turyshev}} \bibnamefont{and}
  \bibinfo{author}{\bibfnamefont{V.~T.} \bibnamefont{Toth}},
  \bibinfo{journal}{Phys. Rev. D} \textbf{\bibinfo{volume}{99}},
  \bibinfo{pages}{024044} (\bibinfo{year}{2019}), \eprint{arXiv:1810.06627
  [gr-qc]}.

\bibitem[{\citenamefont{{Turyshev} and
  {Toth}}(2019)}]{Turyshev-Toth:2019-fin-difract}
\bibinfo{author}{\bibfnamefont{S.~G.} \bibnamefont{{Turyshev}}}
  \bibnamefont{and} \bibinfo{author}{\bibfnamefont{V.~T.}
  \bibnamefont{{Toth}}}, \bibinfo{journal}{Phys. Rev. D}
  \textbf{\bibinfo{volume}{100}}, \bibinfo{pages}{084018}
  (\bibinfo{year}{2019}), \bibinfo{note}{arXiv:1908.01948 [gr-qc]}.

\bibitem[{\citenamefont{{Turyshev} and
  {Toth}}(2020{\natexlab{a}})}]{Turyshev-Toth:2020-image}
\bibinfo{author}{\bibfnamefont{S.~G.} \bibnamefont{{Turyshev}}}
  \bibnamefont{and} \bibinfo{author}{\bibfnamefont{V.~T.}
  \bibnamefont{{Toth}}}, \bibinfo{journal}{Phys. Rev. D}
  \textbf{\bibinfo{volume}{101}}, \bibinfo{pages}{044048}
  (\bibinfo{year}{2020}{\natexlab{a}}), \bibinfo{note}{arXiv:1911.03260
  [gr-qc]}.

\bibitem[{\citenamefont{{Turyshev} and
  {Toth}}(2020{\natexlab{b}})}]{Turyshev-Toth:2020-extend}
\bibinfo{author}{\bibfnamefont{S.~G.} \bibnamefont{{Turyshev}}}
  \bibnamefont{and} \bibinfo{author}{\bibfnamefont{V.~T.}
  \bibnamefont{{Toth}}}, \bibinfo{journal}{Phys. Rev. D}
  \textbf{\bibinfo{volume}{102}}, \bibinfo{pages}{024038}
  (\bibinfo{year}{2020}{\natexlab{b}}), \bibinfo{note}{arXiv:2002.06492
  [astro-ph.IM]}.

\bibitem[{\citenamefont{{Toth} and {Turyshev}}(2020)}]{Toth-Turyshev:2020}
\bibinfo{author}{\bibfnamefont{V.~T.} \bibnamefont{{Toth}}} \bibnamefont{and}
  \bibinfo{author}{\bibfnamefont{S.~G.} \bibnamefont{{Turyshev}}},
  \bibinfo{journal}{Phys. Rev. D}  (\bibinfo{year}{2020}),
  \bibinfo{note}{arXiv:2012.05477 [gr-qc]}.

\bibitem[{\citenamefont{{Abramowitz} and
  {Stegun}}(1965)}]{Abramovitz-Stegun:1965}
\bibinfo{author}{\bibfnamefont{M.}~\bibnamefont{{Abramowitz}}}
  \bibnamefont{and} \bibinfo{author}{\bibfnamefont{I.~A.}
  \bibnamefont{{Stegun}}}, \emph{\bibinfo{title}{Handbook of Mathematical
  Functions: with Formulas, Graphs, and Mathematical Tables.}}
  (\bibinfo{publisher}{Dover Publications, New York; revised edition},
  \bibinfo{year}{1965}).

\bibitem[{\citenamefont{{Park} et~al.}(2017)\citenamefont{{Park}, {Folkner},
  {Konopliv}, {Williams}, {Smith}, and {Zuber}}}]{Park-etal:2017}
\bibinfo{author}{\bibfnamefont{R.~S.} \bibnamefont{{Park}}},
  \bibinfo{author}{\bibfnamefont{W.~M.} \bibnamefont{{Folkner}}},
  \bibinfo{author}{\bibfnamefont{A.~S.} \bibnamefont{{Konopliv}}},
  \bibinfo{author}{\bibfnamefont{J.~G.} \bibnamefont{{Williams}}},
  \bibinfo{author}{\bibfnamefont{D.~E.} \bibnamefont{{Smith}}},
  \bibnamefont{and} \bibinfo{author}{\bibfnamefont{M.~T.}
  \bibnamefont{{Zuber}}}, \bibinfo{journal}{Astron. J.}
  \textbf{\bibinfo{volume}{153}}, \bibinfo{pages}{121} (\bibinfo{year}{2017}).

\bibitem[{\citenamefont{{Roxburgh}}(2001)}]{Roxburgh:2001}
\bibinfo{author}{\bibfnamefont{I.~W.} \bibnamefont{{Roxburgh}}},
  \bibinfo{journal}{Astron. Astrophys.} \textbf{\bibinfo{volume}{377}},
  \bibinfo{pages}{688} (\bibinfo{year}{2001}).

\bibitem[{\citenamefont{{Yates}}(1952)}]{Yates:1952}
\bibinfo{author}{\bibfnamefont{R.~C.} \bibnamefont{{Yates}}},
  \emph{\bibinfo{title}{Epi- and Hypo-Cycloids. A Handbook on Curves and Their
  Properties}} (\bibinfo{publisher}{J. W. Edwards: Ann Arbor, MI},
  \bibinfo{year}{1952}).

\bibitem[{\citenamefont{{Stamnes} and
  {Spjelkavik}}(1983)}]{Stamnes-Spjelkavik:1983}
\bibinfo{author}{\bibfnamefont{J.~J.} \bibnamefont{{Stamnes}}}
  \bibnamefont{and}
  \bibinfo{author}{\bibfnamefont{B.}~\bibnamefont{{Spjelkavik}}},
  \bibinfo{journal}{Opt. Acta: Int. J. Opt.} \textbf{\bibinfo{volume}{30}},
  \bibinfo{pages}{1331} (\bibinfo{year}{1983}).

\bibitem[{\citenamefont{{Ohanian}}(1983)}]{Ohanian:1983}
\bibinfo{author}{\bibfnamefont{H.~C.} \bibnamefont{{Ohanian}}},
  \bibinfo{journal}{Astrophys. J.} \textbf{\bibinfo{volume}{271}},
  \bibinfo{pages}{551} (\bibinfo{year}{1983}).

\bibitem[{\citenamefont{{Berry} and {Howls}}(2010)}]{Berry-Howles:2010}
\bibinfo{author}{\bibfnamefont{M.~V.} \bibnamefont{{Berry}}} \bibnamefont{and}
  \bibinfo{author}{\bibfnamefont{C.~J.} \bibnamefont{{Howls}}}, in
  \emph{\bibinfo{booktitle}{NIST Handbook of Mathematical Functions}}, edited
  by \bibinfo{editor}{\bibfnamefont{F.~W.} \bibnamefont{{Olver}}},
  \bibinfo{editor}{\bibfnamefont{D.~W.} \bibnamefont{{Lozier}}},
  \bibinfo{editor}{\bibfnamefont{R.~F.} \bibnamefont{{Boisvert}}},
  \bibnamefont{and} \bibinfo{editor}{\bibfnamefont{C.~W.}
  \bibnamefont{{Clark}}} (\bibinfo{publisher}{Cambridge University Press},
  \bibinfo{address}{Cambridge, UK}, \bibinfo{year}{2010}), pp.
  \bibinfo{pages}{775--793}.

\bibitem[{\citenamefont{Kofler and Arnold}(2006)}]{Kofler:2006}
\bibinfo{author}{\bibfnamefont{J.}~\bibnamefont{Kofler}} \bibnamefont{and}
  \bibinfo{author}{\bibfnamefont{N.}~\bibnamefont{Arnold}},
  \bibinfo{journal}{Phys. Rev. B} \textbf{\bibinfo{volume}{73}}
  (\bibinfo{year}{2006}).

\bibitem[{\citenamefont{{Pearcey}}(1946)}]{Pearcey:1946}
\bibinfo{author}{\bibfnamefont{T.}~\bibnamefont{{Pearcey}}},
  \bibinfo{journal}{Philos. Mag.} \textbf{\bibinfo{volume}{37}},
  \bibinfo{pages}{311} (\bibinfo{year}{1946}).

\bibitem[{\citenamefont{{Connor} and {Farrelly}}(1981)}]{Connor-Farrelly:1981}
\bibinfo{author}{\bibfnamefont{J.~N.~L.} \bibnamefont{{Connor}}}
  \bibnamefont{and}
  \bibinfo{author}{\bibfnamefont{D.}~\bibnamefont{{Farrelly}}},
  \bibinfo{journal}{J. Chem. Phys.} \textbf{\bibinfo{volume}{75}},
  \bibinfo{pages}{2831} (\bibinfo{year}{1981}).

\bibitem[{\citenamefont{{Turyshev} et~al.}(2020)\citenamefont{{Turyshev},
  {Shao}, {Toth}, and et~al.}}]{Turyshev-etal:2020-PhaseII}
\bibinfo{author}{\bibfnamefont{S.~G.} \bibnamefont{{Turyshev}}},
  \bibinfo{author}{\bibfnamefont{M.}~\bibnamefont{{Shao}}},
  \bibinfo{author}{\bibfnamefont{V.~T.} \bibnamefont{{Toth}}},
  \bibnamefont{and} \bibinfo{author}{\bibnamefont{et~al.}},
  \emph{\bibinfo{title}{Direct multipixel imaging and spectroscopy of an
  exoplanet with a solar gravity lens mission}} (\bibinfo{year}{2020}),
  \bibinfo{note}{arXiv:1908.01948 [gr-qc]}.

\bibitem[{\citenamefont{{Barkana}}(1998)}]{Barkana:1998}
\bibinfo{author}{\bibfnamefont{R.}~\bibnamefont{{Barkana}}},
  \bibinfo{journal}{Astrophys. J.} \textbf{\bibinfo{volume}{502}},
  \bibinfo{pages}{531} (\bibinfo{year}{1998}).

\end{thebibliography}

\appendix
\section{Light at the astroid caustic}
\label{sec:appA}

\subsection{Light near the optical axis}
\label{sec:light-caust}

We study the properties of the SGL PSF. For illustrative purposes, we consider only the quadrupole term in (\ref{eq:B2}), which, in this case, may be given in the following compact form:
{}
\begin{eqnarray}
B(\rho, \phi)&=&\frac{1}{{2\pi}}\int_0^{2\pi} d\phi_\xi \exp\Big[-i\Big(\alpha\rho\cos(\phi_\xi-\phi)+\beta_2\cos[2(\phi_\xi-\phi_s)]\Big)\Big],
\label{eq:zer*1=}
\end{eqnarray}
where, for convenience, we used the quantities defined in (\ref{eq:zerJ})   as
{}
\begin{eqnarray}
\alpha&=&k\sqrt\frac{2r_g}{r},~~~~\beta_2=kr_g  J_2\Big(\frac{R_\odot }{\sqrt{2r_gr}}\Big)^2\sin^2\beta_s.
\label{eq:zerJ=}
\end{eqnarray}

To evaluate the integral (\ref{eq:zer*1=}), we use the Jacobi--Anger expansion \cite{Abramovitz-Stegun:1965}:
{}
\begin{eqnarray}
e^{\pm i z\cos \phi}&=&J_0(z)+2\sum_{n=1}^\infty \Big(J_{2n}(z)\cos 2 n \phi \pm i J_{2n-1}(z)\sin (2n-1) \phi \Big).
\label{eq:Ja-Ang}
\end{eqnarray}

This expansion allows one to present the $\beta_2$-term (or the quadrupole term) in (\ref{eq:zer*1=}) as
{}
\begin{eqnarray}
e^{-i \beta_2\cos [2(\phi_\xi-\phi_s)]}&=&J_0(\beta_2)+2\sum_{n=1}^\infty \Big(J_{2n}(\beta_2)\cos [4n (\phi_\xi-\phi_s)] - i J_{2n-1}(\beta_2)\sin [2(2n-1)(\phi_\xi-\phi_s)]  \Big).
\label{eq:Ja-A-beta}
\end{eqnarray}
Using this expression in (\ref{eq:zer*1=}), we have
{}
\begin{eqnarray}
B(\rho, \phi)&=&\frac{1}{{2\pi}}\int_0^{2\pi} d\phi_\xi \exp\Big[-i\Big(\alpha\rho\cos(\phi_\xi-\phi)+\beta_2\cos[2(\phi_\xi-\phi_s)]\Big)\Big]=\nonumber\\
&&\hskip-60pt =\, \frac{1}{{2\pi}}\int_0^{2\pi} d\phi_\xi e^{-i\alpha\rho\cos(\phi_\xi-\phi)}\Big(J_0(\beta_2)+2\sum_{n=1}^\infty \Big(J_{2n}(\beta_2)\cos [4n (\phi_\xi-\phi_s)] - i J_{2n-1}(\beta_2)\sin [2(2n-1)(\phi_\xi-\phi_s)]  \Big)\Big).~~~
\label{eq:zer*beta}
\end{eqnarray}
We now recognize that
{}
\begin{eqnarray}
\frac{1}{{2\pi}}\int_0^{2\pi} d\phi_\xi e^{-i\alpha\rho\cos(\phi_\xi-\phi)}&=& J_0(\alpha\rho),\nonumber\\
\frac{1}{{2\pi}}\int_0^{2\pi} d\phi_\xi e^{-i\alpha\rho\cos(\phi_\xi-\phi)}\cos [4n (\phi_\xi-\phi_s)]&=& J_{4n}(\alpha\rho)\cos [4n (\phi-\phi_s)],
\nonumber\\
\frac{1}{{2\pi}}\int_0^{2\pi} d\phi_\xi e^{-i\alpha\rho\cos(\phi_\xi-\phi)}\sin [2(2n-1)(\phi_\xi-\phi_s)]&=& -J_{2(2n-1)}(\alpha\rho)\sin [2(2n-1) (\phi-\phi_s)],
\label{eq:zer*beta1}
\end{eqnarray}
and rewrite (\ref{eq:zer*beta}) as
{}
\begin{eqnarray}
B(\rho, \phi)&=&\frac{1}{{2\pi}}\int_0^{2\pi} d\phi_\xi \exp\Big[-i\Big(\alpha\rho\cos(\phi_\xi-\phi)+\beta\cos[2(\phi_\xi-\phi_s)]\Big)\Big]=\nonumber\\
&&\hskip-60pt =\, J_0(\alpha\rho)J_0(\beta_2)+2\sum_{n=1}^\infty \Big(J_{2n}(\beta_2)J_{4n}(\alpha\rho)\cos [4n (\phi-\phi_s)] + i J_{2n-1}(\beta_2)J_{2(2n-1)}(\alpha\rho)\sin [2(2n-1)(\phi-\phi_s)]  \Big).
\label{eq:zer*beta2}
\end{eqnarray}

Therefore, $B(\rho, \phi)$ has the form
{}
\begin{eqnarray}
B(\rho, \phi)&=&\Big(J_0(\alpha\rho)J_0(\beta_2)+2\sum_{n=1}^\infty J_{2n}(\beta_2)J_{4n}(\alpha\rho)\cos [4n (\phi-\phi_s)]\Big) + \nonumber\\
&&\hskip 60pt +\,i\Big(2\sum_{n=1}^\infty J_{2n-1}(\beta_2)J_{2(2n-1)}(\alpha\rho)\sin [2(2n-1)(\phi-\phi_s)]  \Big).
\label{eq:zer*beta3}
\end{eqnarray}

To derive the PSF, we need to square the expression (\ref{eq:zer*beta3}), which results in
{}
\begin{eqnarray}
\hskip -10pt
B^2(\rho, \phi)&=&\Big(J_0(\alpha\rho)J_0(\beta_2)+2\sum_{n=1}^\infty J_{2n}(\beta_2)J_{4n}(\alpha\rho)\cos [4n (\phi-\phi_s)]\Big)^2 +
\nonumber\\&&\hskip 67pt +\,
\Big(2\sum_{n=1}^\infty J_{2n-1}(\beta_2)J_{2(2n-1)}(\alpha\rho)\sin [2(2n-1)(\phi-\phi_s)]  \Big)^2.~~~~
\label{eq:zer*beta4}
\end{eqnarray}

Given the properties of the Bessel functions \cite{Abramovitz-Stegun:1965}, we can see that as we approach the optical axis, $\rho\rightarrow0$ the terms in the expression  (\ref{eq:zer*beta4}) behave as $J_0(\alpha\rho)\rightarrow1$  and $J_n(\alpha\rho)\rightarrow0$, resulting in the following asymptotic behavior:
{}
\begin{eqnarray}
\lim_{\rho\rightarrow 0}
B^2(\rho, \phi)&=&\lim_{\rho\rightarrow 0} \Big\{J^2_0(\alpha\rho)J^2_0(\beta_2)+{\cal O}\Big(J_n^2(\alpha\rho)\Big)\Big\}= J^2_0(\beta_2)\simeq\frac{2}{\pi\beta_2}\cos^2\big(\beta_2-{\textstyle\frac{\pi}{4}}\big).~~~~
\label{eq:j2-lim}
\end{eqnarray}
The PSF of the SGL in the spherically symmetric case (i.e., monopole) was given by $J^2(\alpha\rho)$, which in the limit $\rho\to 0$ yielded 1. As seen from the result (\ref{eq:j2-lim}), In the case of a non-negligible quadrupole  contribution the intensity of the EM field on the optical axis is adjusted by a factor of  ${2}/{\pi\beta_2}$. In the case when other zonal harmonics are present, this value is further reduced by $J^2_0(\alpha\rho)J^2_0(\beta_2)..J^2_0(\beta_{2n})$, where $\beta_{2n}$ are the multipole terms defined similarly to that of a quadrupole in (\ref{eq:zerJ=}). Depending on the value of a particular multipole, this behavior may be presented as $({2}/{\pi\beta_2})\cos^2(\beta_2-{\textstyle\frac{\pi}{4}})({2}/{\pi\beta_4})\cos^2(\beta_4-{\textstyle\frac{\pi}{4}})...({2}/{\pi\beta_{2n}})\cos^2(\beta_{2n}-{\textstyle\frac{\pi}{4}})$. Thus, the light intensity on the optical axis is significantly reduced by the presence of multipole caustics, with the majority of light being deposited in the interior region of the caustic and, more notably, near the caustic boundary.

\subsection{Light at the cusp}
\label{sec:light-cusp}

Using the Jacobi--Anger expansion \cite{Abramovitz-Stegun:1965} that was discussed in the preceding section, we may evaluate (\ref{eq:zer*1=}) directly at a cusp of the astroid caustic. As we discussed earlier, the four cusps form when $(\phi-\phi_s)=k{\textstyle\frac{\pi}{2}}$. For these cusps, (\ref{eq:zer*beta3}) takes the form
{}
\begin{eqnarray}
B(\rho, \phi)\Big|_{\phi-\phi_s=k{\textstyle\frac{\pi}{2}}}&=&\Big(J_0(\alpha\rho)J_0(\beta_2)+2\sum_{n=1}^\infty J_{2n}(\beta_2)J_{4n}(\alpha\rho)\Big).
\label{eq:zer*beta3-cusp}
\end{eqnarray}

We note that along the directions towards the four cusps, the complex amplitude $B(\rho, \phi)$ is a real-valued function.  One important outcome from this observation is that there are no wavelength-dependent phase terms and, thus, not chromatic effects along these directions, which are characterized by constructive interference.

Based on (\ref{eq:q_caust_j1})--(\ref{eq:q_caust_j2}), the cusp is reached when $\alpha\rho=4\beta_2$, yielding the following expression for the PSF at any of the four cusps of the astroid caustic (also from (\ref{eq:zer*beta4})):
{}
\begin{eqnarray}
{\tt PSF}_{\rm cusp}&=&\Big(J_0(\beta_2)J_0(4\beta_2)+2\sum_{n=1}^\infty J_{2n}(\beta_2)J_{4n}(4\beta_2)\Big)^2.
\label{eq:zer*beta3-cusp-psf}
\end{eqnarray}
Although expression (\ref{eq:zer*beta3-cusp-psf}) is rather compact, it is not very convenient for practical use many terms would need to be retained in the sum for useful accuracy.  In fact, the typical number of the required terms is $n\gtrsim\beta_2$.

\subsection{Light in the vicinity of the cusp}
\label{sec:light-cusp-vic}

To evaluate the behavior of the complex amplitude in the regions near the cusps, we take (\ref{eq:zer*1=}) and perform the coordinate transformation to the caustic. For this, without loss of generality, we set $\phi_s=0$. Using parametric equations describing the astroid caustic (\ref{eq:q_caust_j1})--(\ref{eq:q_caust_j2}), we transform (\ref{eq:zer*1=}):
{}
\begin{eqnarray}
B(\phi)&=&\frac{1}{{2\pi}}\int_0^{2\pi} d\phi_\xi \exp\Big[-i\beta_2\Big(\cos(2\phi_\xi)+4\cos(\phi_\xi-\phi)-2\sin2\phi\sin(\phi_\xi+\phi)\Big)\Big],
\label{eq:zer*1=caust}
\end{eqnarray}
where the phase is equivalent to that for the $\beta_2$ part of (\ref{eq:ph*f1}). There is no closed form expression known for all the possible values of $\phi$ along the caustic. Clearly, this integral may be evaluated numerically.  However, we can evaluate this integral analytically in the small vicinity of the cusp.  To demonstrate this,  we consider the cusp at $\phi=0$ and  expand the phase of (\ref{eq:zer*1=caust}) in the small vicinity of $\phi=0$, while treating the angles $\phi\ll1$. Under these conditions, the phase of (\ref{eq:zer*1=caust}) transforms as
{}
\begin{eqnarray}
\varphi(\rho, \phi)&=&-\beta_2\Big(\cos2\phi_\xi+4(1-{\textstyle\frac{3}{2}}\phi^2)\cos\phi_\xi  +{\cal O}(\phi^3) \Big)=
-8\beta_2\Big(\cos^2{\textstyle\frac{1}{2}}\phi_\xi-{\textstyle\frac{3}{4}}\phi^2\Big)^2  +3(1-2\phi^2)\beta_2+{\cal O}(\phi^3),~~~
\label{eq:zer*1=pha}
\end{eqnarray}
resulting in the following form of (\ref{eq:zer*1=caust}):
{}
\begin{eqnarray}
B(\phi)&=&e^{i3(1-2\phi^2)\beta_2}\frac{1}{{2\pi}}\int_0^{2\pi} d\phi_\xi \exp\Big[-i8\beta_2\Big(\cos^2{\textstyle\frac{1}{2}}\phi_\xi-{\textstyle\frac{3}{4}}\phi^2\Big)^2+{\cal O}(\phi^3)\Big].
\label{eq:zer*1=caust2}
\end{eqnarray}
Introducing the new variable ${\textstyle\frac{1}{2}}\phi_\xi=u$, we expand the integrand of (\ref{eq:zer*1=caust2}) in terms of small angle $\phi\ll1$, and obtain the following, valid to the order of ${\cal O}(\phi^3)$:
{}
\begin{eqnarray}
B(\phi)e^{-i3(1-2\phi^2)\beta_2}&=&\frac{1}{{\pi}}\int_0^{\pi} d u \exp\Big[-i8\beta_2\Big(\cos^2 u-{\textstyle\frac{3}{4}}\phi^2\Big)^2\Big]=\nonumber\\
&&\hskip-80pt=\, {\tt PFQ}\Big[\big\{{\textstyle\frac{1}{8}},{\textstyle\frac{3}{8}},{\textstyle\frac{5}{8}},{\textstyle\frac{7}{8}}\big\},\big\{{\textstyle\frac{1}{4}},{\textstyle\frac{1}{2}},{\textstyle\frac{1}{2}},{\textstyle\frac{3}{4}},1\big\},-16\beta_2\Big]-
3i\beta_2 \,
{\tt PFQ}\Big[\big\{{\textstyle\frac{5}{8}},{\textstyle\frac{7}{8}},{\textstyle\frac{9}{8}},{\textstyle\frac{11}{8}}\big\},\big\{{\textstyle\frac{3}{4}},1,{\textstyle\frac{5}{4}},{\textstyle\frac{3}{2}},{\textstyle\frac{3}{2}}\big\},-16\beta_2\Big]+\nonumber\\
&&\hskip-106pt
+\,6\beta_2\Big\{
5\beta_2 \,
{\tt PFQ}\Big[\big\{{\textstyle\frac{7}{8}},{\textstyle\frac{9}{8}},{\textstyle\frac{11}{8}},{\textstyle\frac{13}{8}}\big\},\big\{1,{\textstyle\frac{5}{4}},{\textstyle\frac{3}{2}},{\textstyle\frac{3}{2}},{\textstyle\frac{7}{4}}\big\},-16\beta_2\Big]+i\, {\tt PFQ}\Big[\big\{{\textstyle\frac{3}{8}},{\textstyle\frac{5}{8}},{\textstyle\frac{7}{8}},{\textstyle\frac{9}{8}}\big\},\big\{{\textstyle\frac{1}{2}},{\textstyle\frac{1}{2}},{\textstyle\frac{3}{4}},1,{\textstyle\frac{5}{4}}\big\},-16\beta_2\Big]\Big\}\phi^2+{\cal O}(\phi^3).~~~
\label{eq:zer*1=caust3}
\end{eqnarray}
where ${\tt PFQ}[\{p\},\{q\},-x]$ is the hypergeometric PFQ function \cite{Abramovitz-Stegun:1965}.
In the case of $\beta_2\gg1$, this expression may be given in its asymptotic form:
{}
\begin{eqnarray}
B(\phi)e^{-i3(1-2\phi^2)\beta_2}&=& \frac{1}{2\sqrt{\pi\beta_2}} \cos\Big(8\beta_2-{\textstyle\frac{\pi}{4}}\Big)+\nonumber\\
&+&
\sqrt{\frac{\pi}{2}}\frac{1}{32\beta_2^\frac{3}{4}}
\Gamma[-{\textstyle\frac{1}{4}}]\Gamma[{\textstyle\frac{1}{4}}]\frac{8\Gamma[{\textstyle\frac{3}{8}}]\Gamma[{\textstyle\frac{5}{8}}]+i\Gamma[-{\textstyle\frac{1}{8}}]\Gamma[{\textstyle\frac{1}{8}}]}{\Gamma[-{\textstyle\frac{1}{8}}]\Gamma^2[{\textstyle\frac{1}{8}}]\Gamma[{\textstyle\frac{3}{8}}]\Gamma^2[{\textstyle\frac{5}{8}}]}+
\sqrt{\frac{\pi}{2}}\frac{1}{2\beta_2^\frac{1}{4}}
\Gamma[{\textstyle\frac{1}{4}}]\Gamma[{\textstyle\frac{3}{4}}]\frac{\Gamma[{\textstyle\frac{1}{8}}]\Gamma[{\textstyle\frac{7}{8}}]-i\Gamma[-{\textstyle\frac{3}{8}}]\Gamma[{\textstyle\frac{5}{8}}]}{\Gamma[{\textstyle\frac{1}{8}}]\Gamma^2[{\textstyle\frac{3}{8}}]\Gamma[{\textstyle\frac{5}{8}}]\Gamma^2[{\textstyle\frac{7}{8}}]}
+\nonumber\\
&&\hskip-80pt
+\,\bigg\{3\Big(-\frac{1}{128}\frac{1}{\sqrt{\pi\beta_2}}+i \sqrt{\frac{\beta_2}{\pi}}\Big) \Big(\cos\big(8\beta_2-{\textstyle\frac{\pi}{4}}\big)-i\sin\big(8\beta_2-{\textstyle\frac{\pi}{4}}\big)\Big)
+\nonumber\\
&&\hskip-40pt
+\,
\sqrt{\frac{\pi}{2}}\frac{3}{16\beta_2^\frac{3}{4}}
\Gamma[-{\textstyle\frac{1}{4}}]\Gamma[{\textstyle\frac{1}{4}}]\frac{3\Gamma[-{\textstyle\frac{3}{8}}]\Gamma[{\textstyle\frac{3}{8}}]-i\Gamma[-{\textstyle\frac{1}{8}}]\Gamma[{\textstyle\frac{1}{8}}]}{\Gamma^2[-{\textstyle\frac{3}{8}}]\Gamma[-{\textstyle\frac{1}{8}}]\Gamma^2[{\textstyle\frac{1}{8}}]\Gamma[{\textstyle\frac{3}{8}}]}+
\sqrt{\frac{\pi}{2}}\frac{3}{16\beta_2^\frac{1}{4}}
\Gamma[-{\textstyle\frac{1}{4}}]\Gamma[{\textstyle\frac{1}{4}}]\frac{\Gamma[-{\textstyle\frac{1}{8}}]\Gamma[{\textstyle\frac{1}{8}}]+8i\Gamma[-{\textstyle\frac{3}{8}}]\Gamma[{\textstyle\frac{5}{8}}]}{\Gamma^2[-{\textstyle\frac{1}{8}}]\Gamma[{\textstyle\frac{1}{8}}]\Gamma^2[{\textstyle\frac{3}{8}}]\Gamma^2[{\textstyle\frac{5}{8}}]}
\nonumber\\
&+&
\sqrt{\frac{\pi}{2}}3 \beta_2^\frac{1}{4}
\Gamma[{\textstyle\frac{1}{4}}]\Gamma[{\textstyle\frac{3}{4}}]\frac{\Gamma[{\textstyle\frac{1}{8}}]\Gamma[{\textstyle\frac{7}{8}}]+i\Gamma[-{\textstyle\frac{3}{8}}]\Gamma[{\textstyle\frac{5}{8}}]}{\Gamma^2[{\textstyle\frac{1}{8}}]\Gamma[{\textstyle\frac{3}{8}}]\Gamma^2[{\textstyle\frac{5}{8}}]\Gamma[{\textstyle\frac{7}{8}}]}\bigg\}\phi^2+{\cal O}(\phi^3).~~~
\label{eq:zer*1=caust3-tre}
\end{eqnarray}

Using this result, one may evaluate the magnitude of the PSF at the cusp of the caustic:

{}
\begin{eqnarray}
{\tt PSF}_{\rm cusp}(\phi)&\simeq&\frac{0.118}{\sqrt{\beta_2}}+ \Big(\frac{0.048}{\beta_2^\frac{3}{4}}-\frac{0.03}{\beta_2^\frac{5}{4}}\Big)\cos \big(8\beta_2-{\textstyle\frac{3\pi}{8}}\big)+
\frac{0.03}{\beta_2}
- \Big\{2\,\beta_2^\frac{1}{4}\, \cos \big(8\beta_2-{\textstyle\frac{\pi}{8}}\big)+ 0.05 \Big\}\phi^2+{\cal O}(\phi^3).~~~
\label{eq:zer*1=caust3-num}
\end{eqnarray}

As a result, we see that at the cusp, the PSF reaches its largest value on the caustic and then it decreases on both sizes of the cusp.  (Note that the PSF reaches its maximum not on the caustic boundary but inside, as discussed in Sec.~\ref{sec:wo-cusp}.) The presence of the $\phi$-dependent term then rapidly suppresses the PSF its peak value.

\end{document}